\documentclass[%
 reprint,
superscriptaddress,
showpacs,
 amsmath,amssymb,
 aps,
floatfix,
]{revtex4-1}

\usepackage{graphicx}
\usepackage{dcolumn}
\usepackage{bm}

\usepackage{float}
\usepackage{amsfonts}
\usepackage{latexsym}
\usepackage{amssymb}
\usepackage{mathrsfs}
\usepackage{amsmath}
\usepackage{url} 
\usepackage{color}

\makeatletter
\newcommand*{\bigcdot}{}
\DeclareRobustCommand*{\bigcdot}{%
  \mathbin{\mathpalette\bigcdot@{}}%
}
\newcommand*{\bigcdot@scalefactor}{.5}
\newcommand*{\bigcdot@widthfactor}{1.15}
\newcommand*{\bigcdot@}[2]{%
  \sbox0{$#1\vcenter{}$}
  \sbox2{$#1\cdot\m@th$}%
  \hbox to \bigcdot@widthfactor\wd2{%
    \hfil
    \raise\ht0\hbox{%
      \scalebox{\bigcdot@scalefactor}{%
        \lower\ht0\hbox{$#1\bullet\m@th$}%
      }%
    }%
    \hfil
  }%
}
\makeatother

\begin{document}

\preprint{APS/123-QED}
\title{A Framework for the Time- and Frequency-Domain Assessment of High-Order Interactions in Brain and Physiological Networks}


\author{Luca Faes}
\email{luca.faes@unipa.it}
\affiliation{%
 Department of Engineering, University of Palermo, Italy
}

\author{Gorana Mijatovic}
\affiliation{
 Faculty of Technical Sciences, University of Novi Sad, Serbia}

\author{Yuri Antonacci}
\affiliation{
Department of Physics and Chemistry "Emilio Segrè", University of Palermo, Italy
}

\author{Riccardo Pernice}
\affiliation{
Department of Engineering, University of Palermo, Italy
}

\author{Chiara Barà}
\affiliation{
Department of Engineering, University of Palermo, Italy
}

\author{Laura Sparacino}
\affiliation{
Department of Engineering, University of Palermo, Italy
}

\author{Marco Sammartino}
\affiliation{
Department of Engineering, University of Palermo, Italy
}

\author{Alberto Porta}
\affiliation{
Department of Biomedical Sciences for Health, University of Milano, Italy, and  Department of Cardiothoracic, Vascular Anesthesia and Intensive Care, IRCCS Policlinico San Donato, Italy
}

\author{Daniele Marinazzo}
\affiliation{
Department of Data Analysis, University of Ghent, Belgium
}

\author{Sebastiano Stramaglia}
\affiliation{
Department of Physics, University of Bari Aldo Moro, and INFN Sezione di Bari, Italy
}


\date{\today}

\date{\today}

\begin{abstract}
While the standard network description of complex systems is based on quantifying the link between pairs of system units, higher-order interactions (HOIs) involving three or more units often play a major role in governing the collective network behavior.
This work introduces an approach to quantify pairwise and HOIs for multivariate rhythmic processes interacting across multiple time scales.
We define the so-called O-information rate (OIR) as a new metric to assess HOIs for multivariate time series, and propose a framework to decompose the OIR into measures quantifying Granger-causal and instantaneous influences, as well as to expand all measures in the frequency domain. The framework exploits the spectral representation of vector autoregressive and state-space models to assess the synergistic and redundant interaction among groups of processes, both in specific bands of biological interest and in the time domain after whole-band integration.
Validation of the framework on simulated networks illustrates how the spectral OIR can highlight redundant and synergistic HOIs emerging at specific frequencies, which cannot be detected using time-domain measures. The applications to physiological networks described by heart period, arterial pressure and respiration variability measured in healthy subjects during a protocol of paced breathing, and to brain networks described by electrocorticographic signals acquired in an animal experiment during anesthesia, document the capability of our approach to identify informational circuits relevant to well-defined cardiovascular oscillations and brain rhythms and related to specific physiological mechanisms involving autonomic control and altered consciousness. The proposed framework allows a hierarchically-organized evaluation of time- and frequency-domain interactions in dynamic networks mapped by multivariate time series, and its high flexibility and scalability make it suitable for the investigation of networks beyond pairwise interactions in neuroscience, physiology and many other fields.
\end{abstract}

\maketitle

\section{Introduction}
\label{sec_intro}
The increasing availability of large-scale and fine-grained recordings of biomedical signals and physiological time series is nowadays boosting the development of new methods for the data-driven modelling of complex biological systems. Among them, the network representation of physiological systems is probably the most used approach to the description of multivariate time series measured from these systems \cite{barabasi2013network}. Paradigmatic instances of this approach come from the neurosciences, where the organizational principles of functional segregation and integration in the brain are typically studied through the theoretical and empirical tools of Network Neuroscience \cite{bassett2017network}, and from integrative physiology, where the reductionist approach of studying  the function of an organ system in isolation is complemented by the holistic investigation of collective interactions among diverse systems performed in the field of Network Physiology \cite{bashan2012network}.

Data-driven methods for the inference and analysis of physiological networks are based on building a network model out of a set of observed time series, in which nodes represent the units composing the observed system (being, e.g., distinct neural populations or organ systems) and connecting edges map functional dependencies between pairs of units (descriptive, e.g., of brain connectivity or cardiovascular interactions) \cite{rubinov2010complex,lehnertz2020human}.  
Nevertheless, in spite of the ubiquitous utilization of pairwise measures to describe interactions in a network, there is mounting evidence that such measures cannot full capture the interplay among the multiple units of a complex system \cite{battiston2020networks}. 
In fact, brain and physiological networks -among others- exhibit collective behaviors which are integrated at different hierarchical levels, thus displaying interactions that involve more than two network nodes. These so-called \textit{high-order interactions} (HOIs) occur for instance when brain dynamics require the joint examination of multiple units to be predicted accurately \cite{stramaglia2012expanding,stramaglia2016synergetic}, or when cardiovascular interactions are influenced by the effects of the respiratory activity \cite{porta2011accounting,porta2017quantifying}.

The recognized need to study networks beyond the framework of pairwise interactions calls for the theoretical definition and practical development of methods to assess HOIs among multiple time series. Various metrics solidly grounded in the general field of information theory have been proposed in recent years for this purpose, all attempting to capture the redundant or synergistic information shared by groups of random variables or processes \cite{lizier2018information,faes2017information,rosas2019quantifying}. In broad terms, synergy arises from statistical interactions that can be found collectively in a network but not in parts of it considered separately, while redundancy refers to group interactions that can be explained by the communication of sub-groups of variables. The most popular measures of synergy and redundancy are those based on the interaction information and partial information decomposition of random variables, also extended to assess directed interactions in dynamic physiological processes \cite{stramaglia2012expanding,porta2017quantifying,lizier2018information,faes2017information}. A recently-proposed measure is the so-called O-information, a metric capable to reveal synergy- and redundancy-dominated interactions in a network of multiple interacting variables \cite{rosas2019quantifying}. Its symmetric nature, the fact that it scales with the network size, and the possibility to compute it for dynamic processes make the O-information a very promising tool for the practical analysis of multivariate physiological dynamics \cite{stramaglia2021quantifying}.

A main limitation of the information-theoretic measures proposed so far to investigate HOIs in network systems is that they characterize the system dynamics with one single value reflecting the aggregate effect of interactions possibly occurring at different time scales. However, the time series measured at the nodes of brain and physiological networks are typically rich of oscillatory content: for instance, cardiovascular and electroencephalographic (EEG) interactions occur through the coupling of rhythms in different frequency bands with different physiological meaning \cite{porta2015wiener,he2019electrophysiological}. Remarkably, the amplitude of oscillations and the coupling strength may vary with frequency, and HOIs can have different nature for different rhythms because synergistic and redundant behaviors may alternate in separate frequency bands \cite{faes2021information,antonacci2021measuring}. Therefore, there is the need to connect the 
spectral representation of information-theoretic measures with the HOI description of complex networks to overcome spectral pairwise approaches \cite{geweke1982measurement,chicharro2011spectral}.
To this end, the present study introduces a new framework for the time- and frequency-domain analysis of HOIs in multivariate stochastic processes mapping the activity of network systems. 
Building on our recent efforts to compute multivariate information measures in the frequency domain \cite{faes2021information,antonacci2021measuring}, we generalize and extend them in many directions. First, we define a new measure, the \textit{O-information rate} (OIR), which generalizes the mutual information rate (MIR) of bivariate processes using the same rationale whereby the O-information generalizes the mutual information (MI) between random variables. Then, we provide both a causal decomposition and a spectral expansion of the OIR, thereby connecting it with well-known and widely used measures of coupling and Granger causality formulated in the time and frequency domains \cite{chicharro2011spectral}. Causal and spectral measures are defined from the vector autoregressive (VAR) formulation of multivariate Gaussian stochastic processes \cite{faes2012measuring}, in a way such that the spectral integration of each frequency domain measure yields the corresponding time domain measure. Further, to allow their closed-form computation, all measures composing the time-and frequency-domain OIR are implemented exploiting the state-space (SS) representation of VAR processes \cite{barnett2015granger}.

In this paper, the proposed framework is first illustrated on theoretical examples of simulated VAR processes featuring HOIs of different type and order. Then, it is tested in two practical applications of of brain and physiological networks where HOIs are expected to play a crucial role in governing collective dynamics: beat-to-beat variability series of heart period, arterial pressure and respiration measured during a protocol of paced breathing \cite{porta2011accounting}, and multi-electrode invasive EEG signals acquired in an animal experiment of altered consciousness \cite{yanagawa2013large}. The time- and frequency-domain measures of bivariate and higher-order interactions provided by the framework are collected in the OIR Matlab toolbox, freely available for download at www.lucafaes.net/OIR.html.

\section{Framework to measure high-order interactions in multivariate processes} \label{sec_methods}
This section presents the framework to measure dynamic interactions among $Q$ stationary stochastic processes $Y=\{Y_1, \ldots, Y_Q\}$, grouped in $M$ blocks $X=\{X_1, \ldots, X_M\}$ which can be thought as descriptive of the activity of a network formed by $M$ dynamic systems (the $i^{th}$ block has dimension $M_i$, so that $Q=\sum_{i=1}^M M_i$).
To highlight the dynamic nature of the process $X_i$, we denote as $X_{i,n}$, $X^k_{i,n}=[X_{i,n-1} \cdots X_{i,n-k}]$, and $X^-_{i,n}=\lim_{k \to \infty}X^k_{i,n}$ the random variables that sample the process  at the present time $n$, over the past $k$ lags, and over the whole past history, respectively.

In the following, interactions are characterized providing definitions of high-order measures as well as of their causal decomposition and spectral expansion, and describing the approach implemented for their computation.
While the subsections are self-explanatory, we refer the reader to the supplemental material for detailed mathematical treatments.

\subsection{O-information rate} \label{ssec_OIR}
We start recalling the concept of \textit{information rate}, which quantifies the time density of the average information in a stochastic process. For a generic process $X_i$, the entropy rate is defined as the conditional entropy of the present state given the past history, i.e. $H_{X_i}=H(X_{i,n}|X_{i,n}^-)$ \cite{cover1991elements}.
If two processes $X_i$ and $X_j$ are considered, the definition of MI and the use of basic information rules \cite{cover1991elements} lead to derive MIR as $I_{X_i;X_j}=H_{X_i}+H_{X_j}-H_{X_i,X_j}$.

While the MIR is a dynamic measure of pairwise interdependence, multivariate measures involving more than two processes can be used to assess HOIs. Here, following recent works \cite{rosas2019quantifying,stramaglia2021quantifying}, we measure the \textit{organizational structure} of a group of stochastic processes introducing the so-called O-information rate (OIR). Specifically, the OIR of $N$ processes taken from the set $\{X_1, \ldots, X_M\}$ is defined via the recursion
\begin{subequations}
\begin{align}
   \Omega_{X^2}&=0, \label{OIR2}\\ 
   \Omega_{X^N}&=\Omega_{X^N_{-j}}+\Delta_{X^N_{-j};X_j}, N \geq 3 \label{OIRN}
\end{align}
\label{OIRrec}
\end{subequations}
\noindent{where $X^N=\{X_{i_1},...,X_{i_N}\}$, $i_1, \ldots ,i_N \in \{1,\ldots M \}, N\leq M$, is the analyzed group of processes, $X^N_{-j}=X^N \backslash X_{j}$ is the subset where $X_{j}$ is removed ($j \in \{i_1,\ldots i_N \}$), and  where the variation of the OIR obtained with the addition of $X_j$ to $X^N_{-j}$ is the quantity}
\begin{equation}
   \Delta_{X^N_{-j};X_j}=(2-N)I_{X_j;X^N_{-j}} + \sum_{\substack{m=1 \\ m \neq j}}^{N}I_{X_j;X^N_{-mj}},
   \label{deltaOIR}
\end{equation}
with $X^N_{-mj}=X^N \backslash \{X_m,X_j\}$.
The OIR $\Omega_{X^N}$ is a symmetric measure capturing the balance between high- and low-order statistical constraints in the dynamic interactions occurring within $X^N$: $\Omega_{X^N}>0$ reflects a dominance of low-order constraints, also known as \textit{redundancy}, while $\Omega_{X^N}<0$ indicates that high-order constraints prevail, denoting \textit{synergy}.
In turn, the sign of the OIR increment defined in (\ref{deltaOIR}) detects the informational character of the circuits which link the $j^{th}$ process with the remaining $N-1$ processes of $X^N$: the information that $X_j$ shares with $X^N_{-j}$ is redundant when $\Delta_{X^N_{-j};X_j}>0$, while it is synergistic when $\Delta_{X^N_{-j};X_j}<0$.
Note that when $N=3$ processes $X^3=\{X_i,X_k,X_j\}$ are considered, substituting (\ref{OIR2}) in (\ref{OIRN}) yields $\Omega_{X^3}=\Delta_{\{X_i,X_k\};X_j}$, which expanded with (\ref{deltaOIR}) gives a dynamic version of the well-known interaction information \cite{mcgill1954multivariate}, i.e. $\Omega_{\{X_i,X_k,X_j\}}=I_{X_i;X_j}+I_{X_k;X_j}-I_{\{X_i,X_k\};X_j}$.

Now we move to provide a causal decomposition of the OIR increment $\Delta_{X^N_{-j};X_j}$. To this end, we note that this measure is obtained inserting $N$ different MIR values in (\ref{deltaOIR}), i.e. the MIRs between the processes $Z_1=X_j$ and $Z_2=X^N_{-mj}$ where $Z_1$ is fixed and $Z_2$ varies with $m=0,1,\ldots,N, m \neq j$ ($Z_2=X^N_{-j}$ when $m=0$). The MIR $I_{Z_1;Z_2}$ can be formulated according to the expansion \cite{chicharro2011spectral}
\begin{equation}
   I_{Z_1;Z_2}=T_{Z_1 \rightarrow Z_2} + T_{Z_2 \rightarrow Z_1} + I_{Z_1 \boldsymbol{\bigcdot} Z_2},
   \label{MIRexpZ}
\end{equation}
where $T_{Z_i \rightarrow Z_j}=I(Z_{j,n};Z_{i,n}^-|Z_{j,n}^-)$ is the transfer entropy (TE) from $Z_i$ to $Z_j$ ($i,j=1,2$), and $I_{Z_1 \bigcdot Z_2}=I(Z_{1,n};Z_{2,n}|Z_{1,n}^-,Z_{2,n}^-)$ represents the instantaneous information shared between $Z_1$ and $Z_2$.
The TE is a well-known measure of directed information transfer between two stochastic processes \cite{schreiber2000measuring}, while the instantaneous transfer is a symmetric measure of information shared at zero lag, quantified after removing the common information with the past states of the processes.
The substitution of (\ref{MIRexpZ}) in (\ref{deltaOIR}) allows to decompose the OIR increment as
\begin{equation}
   \Delta_{X^N_{-j};X_j} = \Delta_{X^N_{-j} \rightarrow X_j} + \Delta_{X_j \rightarrow X^N_{-j}} + \Delta_{X^N_{-j} \bigcdot X_j}
   \label{deltaOIRexp},
\end{equation}
where the terms $\Delta_{X^N_{-j} \rightarrow X_j}$ and $\Delta_{X_j \rightarrow X^N_{-j}}$ are derived from the transfer entropies and quantify the informational character of the directed information transfer from $X^N_{-j}$ to $X_j$ and from $X_j$ to $X^N_{-j}$, and the term $\Delta_{X^N_{-j} \bigcdot X_j}$ is derived from the information shared instantaneously between $X_j$ and $X^N_{-j}$ and quantifies its informational character; the informational character of each term is redundant when the term is positive, and synergistic when the term is negative.

\subsection{Linear parametric formulation} \label{ssec_linear}
This section reports the linear parametric formulation of the OIR decomposition, which exploits the knowledge that this formulation captures all of the entropy differences relevant to the various information measures when the observed processes have a joint Gaussian distribution \cite{barrett2010multivariate}. The linear parametric representation of the original vector $Y$ is provided by the vector autoregressive (VAR) model 
\begin{equation}
   Y_n = \sum_{k=1}^{p}\mathbf{A}_k Y_{n-k} + U_n
   \label{VAR},
\end{equation}
where $p$ is the model order, $Y_n=[Y_{1,n} \cdots Y_{Q,n}]^\intercal$ is a $Q$-dimensional vector collecting the present state of all processes, $\mathbf{A}_k$ is the $Q \times Q$ matrix of the model coefficients relating the present with the past of the processes assessed at lag $k$, and $U_n=[U_{1,n} \cdots U_{Q,n}]^\intercal$ is a vector of $Q$ zero-mean white and uncorrelated noises with $Q \times Q$ positive definite covariance matrix $\mathbf{\Sigma}_U=\mathbb{E}[U_n U_n^\intercal]$.
While the VAR model (\ref{VAR}) provides a global representation of the overall multivariate process, to describe the linear interactions relevant to the subset of processes $Z=\{Z_1,Z_2 \}$ for which the MIR decomposition (\ref{MIRexpZ}) is sought
we need to define a \textit{reduced} VAR model involving only those processes. This reduced model is formulated as
\begin{equation}
   Z_n = \sum_{k=1}^{\infty}\mathbf{B}_k Z_{n-k} + W_n
   \label{VARZ},
\end{equation}
where $Z_n$ and $W_n$ are columns vectors of dimension $R=R_1+R_2$ ($R_1=M_j$ is the dimension of $Z_1=X_j$ and $R_2$ is the dimension of $Z_2=X^N_{-mj}$), and $\mathbf{B}_k$ is an $R \times R$ coefficient matrix. Note that the coefficients $\mathbf{B}_k$ and innovations $W_n$ generally differ from those defined for the overall model, and that the order of the reduced model is typically infinite \cite{barnett2015granger}; for this reason, it is strongly advised to identify the reduced VAR model (\ref{VARZ}) in closed form using state space models, as reported in Sect. \ref{ssec_impl}.

The linear parametric representation (\ref{VARZ}) can be used to perform MIR and OIR decomposition in the frequency domain. To this end, the Fourier transfrom (FT) of (\ref{VARZ}) is taken to derive
\begin{equation}
   Z(\omega) = [\mathbf{I}_R - \sum_{k=1}^{\infty}\mathbf{B}_k e^{-\mathbf{j} \omega k}]^{-1} W(\omega)=\mathbf{H}(\omega)W(\omega)
   \label{VARfreq},
\end{equation}
where $Z(\omega)$ and $W(\omega)$ are the Fourier transforms of $Z_n$ and $W_n$, $\omega \in [-\pi,\pi]$ is the normalized angular frequency ($\omega=2\pi \frac{f}{f_s}$ with $f\in [-\frac{f_s}{2},\frac{f_s}{2}]$, being $f_s$ the sampling frequency of the processes), $\mathbf{j}=\sqrt{-1}$ and $\mathbf{I}_R$ is the $R$-dimensional identity matrix. The $R \times R$ matrix $\mathbf{H}(\omega)$ contains the transfer functions relating the FTs of the innovation processes in $W$ to the FTs of the processes in $Z$.
This matrix, as well as the $R \times R$ power spectral density (PSD) matrix of the process defined as the FT of the autocorrelation function ($\mathbf{S}_Z(\omega)=\mathfrak{F}\{\mathbf{R}_{Z}(k)\},\mathbf{R}_{Z}(k)=\mathbb{E}[Z_n Z_n^\intercal]$), can be factorized in four blocks to evidence the spectral properties related to the internal dynamics of $Z_1$ and $Z_2$ through the $R_i \times R_i$ diagonal blocks $\mathbf{S}_{Z_i}(\omega)$ and $\mathbf{H}_{ii}(\omega)$, or to the causal interactions between $Z_1$ and $Z_2$ through the $R_i \times R_j$ off-diagonal blocks $\mathbf{S}_{Z_iZ_j}(\omega)$ and $\mathbf{H}_{ij}(\omega)$, $i,j\in \{1,2\}$.
Then, using spectral factorization 
to express the PSD of $Z$ as $\mathbf{S}_Z(\omega)=\mathbf{H}(\omega)\mathbf{\Sigma}_{W}\mathbf{H}^*(\omega)$, where $\mathbf{\Sigma}_W=\mathbb{E}[W_n W_n^\intercal]$ and $^*$ stands for conjugate transpose, and expanding this factorization to evidence the PSD of $Z_i$ in terms of internal dynamics and causal interactions from $Z_j$, logarithmic spectral measures of the total coupling between $Z_i$ on $Z_j$ and of the causal coupling from $Z_j$ on $Z_i$ can be computed as \cite{geweke1982measurement}
\begin{equation}
    f_{Z_i;Z_j}(\omega)=\log \frac{|\mathbf{S}_{Z_i}(\omega)||\mathbf{S}_{Z_j}(\omega)|}{|\mathbf{S}_{Z}(\omega)|},
   \label{f12}
\end{equation}
\begin{equation}
    f_{Z_j \rightarrow Z_i}(\omega)=\log \frac{|\mathbf{S}_{Z_i}(\omega)|}{|\mathbf{H}_{ii}(\omega)\mathbf{\Sigma}_{W_i}\mathbf{H}_{ii}^*(\omega)|}.
   \label{fi_j}
\end{equation}
Moreover, a spectral measure $f_{Z_i \boldsymbol{\bigcdot} Z_j}(\omega)$ can be defined subtracting the two causal measures (\ref{fi_j}) from the coupling measure (\ref{f12}) so as to satisfy in the frequency domain a decomposition similar to the time-domain decomposition (\ref{MIRexpZ}), i.e.
\begin{equation}
    f_{Z_1;Z_2}(\omega)=f_{Z_1 \rightarrow Z_2}(\omega) + f_{Z_2 \rightarrow Z_1}(\omega) + f_{Z_1 \boldsymbol{\bigcdot} Z_2}(\omega).
   \label{MIRexpZfreq}
\end{equation}
Importantly, the spectral measures in (\ref{MIRexpZfreq}) and the time-domain measures are tightly linked to the similar measures given in the time domain in (\ref{MIRexpZ}). In fact, it can be shown (see, e.g., \cite{chicharro2011spectral}) that integration over the whole frequency axis of the spectral coupling measure (\ref{f12}) returns, with proper scaling, the MIR between the two processes,
\begin{equation}
I_{Z_1;Z_2} = \dfrac{1}{4\pi} \int_{-\pi}^{\pi} f_{Z_1;Z_2}(\omega) \,\textrm{d}\omega \label{F12_f12},
\end{equation}
and that the same relation holds integrating $f_{Z_1 \rightarrow Z_2}(\omega)$, $f_{Z_2 \rightarrow Z_1}(\omega)$ and $f_{Z_1 \boldsymbol{\bigcdot} Z_2}(\omega)$ to get respectively $T_{Z_1 \rightarrow Z_2}$, $T_{Z_2 \rightarrow Z_1}$, and $I_{Z_1 \boldsymbol{\bigcdot} Z_2}$. This spectral integration property gives to the spectral measures $f_{Z_1;Z_2}(\omega)$ and $f_{Z_1 \rightarrow Z_2}(\omega)$,$f_{Z_2 \rightarrow Z_1}(\omega)$, the information-theoretic meaning of density of information shared between the two processes, or transferred from one process to the other, at the angular frequency $\omega$.
We note that, while the coupling measure is always non-negative, the two causal measures can take negative values at some frequencies if the process $Z$ is not strictly causal (i.e. if the innovation covariance $\mathbf{\Sigma}_W$ is not block-diagonal). On the contrary, the measure $f_{Z_1 \boldsymbol{\bigcdot} Z_2}(\omega)$ can take negative values even for strictly causal processes.

The spectral integration property can be exploited not only to compute the time-domain measures in (\ref{MIRexpZ}) as the average of the spectral measures in (\ref{MIRexpZfreq}), but also to achieve a causal decomposition of the OIR formulated for spectral functions. Indeed, it is easy to show that the frequency-specific OIR increment defined in analogy to (\ref{deltaOIR}) as 
\begin{equation}
   \delta_{X^N_{-j};X_j}(\omega)=(2-N)f_{X_j;X^N_{-j}}(\omega) + \sum_{\substack{m=1 \\ m\neq j}}^{N}f_{X_j;X^N_{-mj}}(\omega)
   \label{deltaOIRfreq},
\end{equation}
satisfies the spectral integration property, i.e. $\Delta_{X^N_{-j};X_j} = (1/{4\pi}) \int_{-\pi}^{\pi} \delta_{X^N_{-j};X_j}(\omega) \,\textrm{d}\omega$, and can also be expanded through a causal decomposition similar to (\ref{deltaOIRexp}) as
\begin{equation}
   \delta_{X^N_{-j};X_j}(\omega) = \delta_{X^N_{-j} \rightarrow X_j}(\omega) + \delta_{X_j \rightarrow X^N_{-j}}(\omega) + \delta_{X^N_{-j} \bigcdot X_j}(\omega)
   \label{deltaOIRexpfreq},
\end{equation}
where the three terms on the r.h.s. of (\ref{deltaOIRexpfreq}) are obtained expanding $f_{X_j;X^N_{-j}}(\omega)$ and $f_{X_j;X^N_{-mj}}(\omega)$ in (\ref{deltaOIRfreq}) according to (\ref{MIRexpZfreq}).
Moreover, the spectral OIR increment (\ref{deltaOIRfreq}) can be used to compute recursively a frequency-domain version of the OIR, in analogy to (\ref{OIRrec}), as
\begin{equation}
   \nu_{X^N}(\omega)=\nu_{X^N_{-j}}(\omega)+\delta_{X^N_{-j};X_j}(\omega)
   \label{OIRrecfreq},
\end{equation}
which again satisfies the spectral integration property, i.e. $\Omega_{X^N} = (1/{4\pi}) \int_{-\pi}^{\pi} \nu_{X^N}(\omega) \,\textrm{d}\omega$.
Therefore, the spectral versions of the HOI measures defined in this section can be meaningfully interpreted as densities of the synergistic/redundant character of the information shared between multiple stochastic processes. As shown in the theoretical examples of Sect. \ref{sec_examples} and practical applications of Sect. \ref{sec_applications}, the evaluation of these measures within specific frequency bands allows to assess the informational character of specific oscillations within circuits of nodes of the analyzed network.

\subsection{Framework Implementation} \label{ssec_impl}
This section reports the time- and frequency-domain computation of the OIR and of the terms of its decomposition performed within the framework of state space (SS) models. The advantage of using SS models is that this class of models is closed under the definition of reduced models, i.e. models which contain only some of the original analyzed processes. In other words, while a reduced VAR model like that formulated in (\ref{VARZ}) is generally of infinite order and thus very difficult to identify from finite-length time series, SS models can be reduced maintaining their form and can be therefore identified keeping high computational reliability.

Here, we follow the SS modeling approach of \cite{barnett2015granger} to compute all the MIR terms needed to derive the OIR (\ref{OIRrec}) and to perform the related causal decomposition (\ref{deltaOIRexp}) and spectral expansion (\ref{deltaOIRexpfreq}).
First, we describe the original process $Y$ obeying the VAR representation (\ref{VAR}) using the SS model
\begin{subequations} \label{ISS}
\begin{align}
		S_{n+1} &= \mathbf{A} S_{n} + \mathbf{K} U_{n}, \label{ISSstate} \\
		Y_n &= \mathbf{C} S_{n} + U_{n}, \label{ISSobs}
\end{align}
\end{subequations}
where $S_n=[Y_{n-1}^\intercal \cdots Y_{n-p}^\intercal]^\intercal$ is the $pQ$-dimensional state process and the SS parameters ($\mathbf{A},\mathbf{C},\mathbf{K},\mathbf{V}$) are given by the matrices $\mathbf{C}=[\mathbf{A}_1\cdots\mathbf{A}_p]$, $\mathbf{A}=[\mathbf{C};\mathbf{I}_{Q(p-1)} \mathbf{0}_{Q(p-1)\times Q}]$, $\mathbf{K}=[\mathbf{I}_Q \mathbf{0}_{Q\times Q(p-1)}]^\intercal$, and $\mathbf{V}=\mathbb{E}[U_n U_n^\intercal]=\mathbf{\Sigma}_U$.
Then, to represent the $R$-dimensional process $Z=\{Z_1,Z_2\}$ formed by taking from $Y$ the subset of processes indexed by the elements of $\mathbf{r}=\{\mathbf{r}_1,\mathbf{r}_2\}\subset \{1,\ldots, Q\}$ (where $\mathbf{r}_i$ contains the $R_i$ indices of $Z_i,i=1,2$), we replace (\ref{VARZ}) with a reduced SS model with state equation (\ref{ISSstate}) and observation equation $Z_n=\mathbf{C}^{(\mathbf{r},:)} S_{n} + W_{n}$.
This model has parameters ($\mathbf{A},\mathbf{C}^{(\mathbf{r},:)},\mathbf{K}\mathbf{V}\mathbf{K}^\intercal,\mathbf{V}^{(\mathbf{r},\mathbf{r})},\mathbf{K}\mathbf{V}^{(:,\mathbf{r})}$), where the superscripts $^{(\mathbf{r},:)}$, $^{(:,\mathbf{r})}$, and $^{(\mathbf{r},\mathbf{r})}$ denote selection of the rows and/or columns with indices $\mathbf{r}$ in a matrix.
To exploit the reduced SS model for the Granger-causal analysis of $Z$ it is necessary to lead its form back to that of (\ref{ISS}), which reads \cite{barnett2015granger}
\begin{subequations} \label{ISSZ}
\begin{align}
		S_{n+1} &= \Tilde{\mathbf{A}} S_{n} + \Tilde{\mathbf{K}} W_{n}, \label{ISSZstate} \\
		Z_n &= \Tilde{\mathbf{C}} S_{n} + W_{n}. \label{ISSZobs}
\end{align}
\end{subequations}
The parameters of the model (\ref{ISSZ}) are ($\Tilde{\mathbf{A}},\Tilde{\mathbf{C}},\Tilde{\mathbf{K}},\Tilde{\mathbf{V}}$), of dimension $pQ \times pQ, R\times pQ, pQ \times R, R \times R$; while the state and observation matrices are easily determined as $\Tilde{\mathbf{A}}=\mathbf{A}$ and
and $\Tilde{\mathbf{C}}=\mathbf{C}^{(\mathbf{r},:)}$, the gain $\Tilde{\mathbf{K}}$ and the reduced innovation covariance $\Tilde{\mathbf{V}}=\mathbb{E}[W_n W_n^\intercal]=\mathbf{\Sigma}_W$ must be obtained by solving a discrete algebraic Riccati equation (DARE) (see refs. \cite{barnett2015granger,faes2017multiscale} for detailed derivations).   

After its identification, the reduced model (\ref{ISSZ}) can be analyzed in the frequency domain to compute the spectral measures described in Sect. \ref{ssec_linear}. To this end, the FT of (\ref{ISSZstate}) is computed to derive the PSD of the state process, $S(\omega)$, which is then substituted in the FT of (\ref{ISSZobs}) to obtain \cite{barnett2015granger}
\begin{equation}
   Z(\omega) = \big(\mathbf{I}_R + \Tilde{\mathbf{C}}[\mathbf{I}_{pQ}-\Tilde{\mathbf{A}}e^{-\mathbf{j} \omega}]^{-1}\Tilde{\mathbf{K}}e^{-\mathbf{j} \omega}\big)W(\omega)=\mathbf{H}(\omega)W(\omega)
   \label{ISSfreq},
\end{equation}
The transfer function matrix $\mathbf{H}(\omega)$ is then used to derive the PSD matrix of the reduced process as $\mathbf{S}_Z(\omega)=\mathbf{H}(\omega)\Tilde{\mathbf{V}}\mathbf{H}^*(\omega)$. All these spectral matrices are finally employed as described in Sect. \ref{ssec_linear} to compute the frequency-domain decomposition of the MIR and OIR measures, and to derive the corresponding time-domain measures through spectral integration.

\begin{figure*} [t]
    \centering
    \includegraphics[scale=0.85]{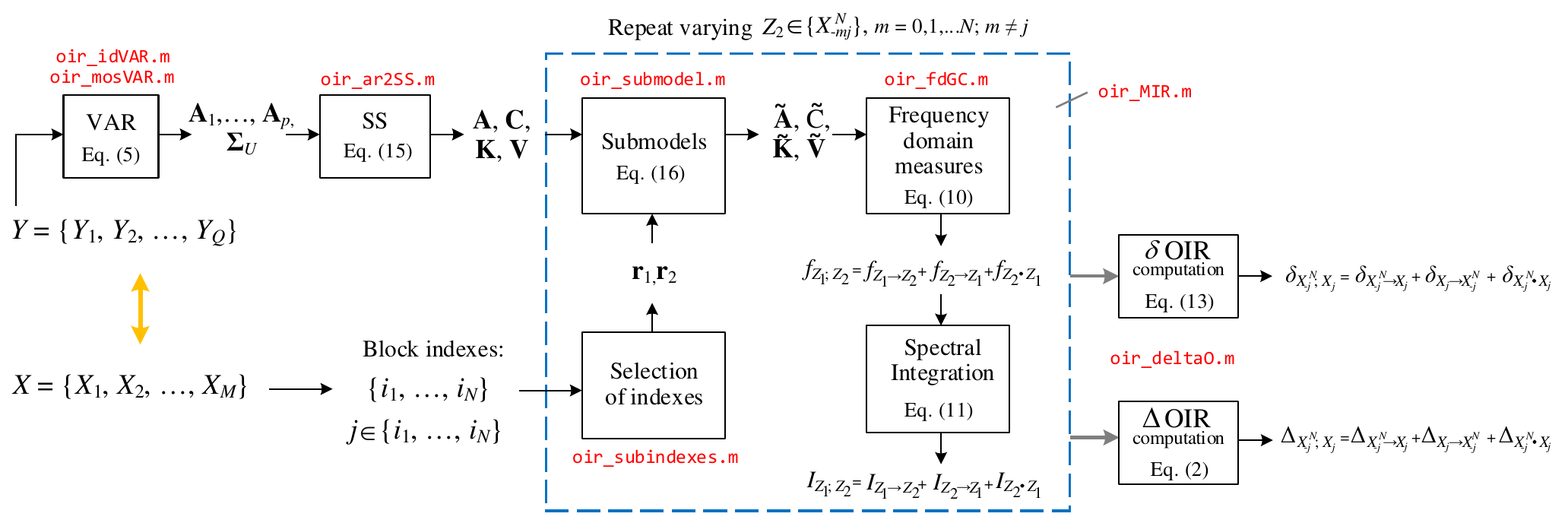}
    \caption{Schematic description of the algorithmic implementation of the framework for OIR computation. Blocks represent the operations applied to the input stochastic processes and/or to their parametric representations for deriving the different time- and frequency-domain interaction measures. The equations and the functions of the OIR Matlab Toolbox implemented for each step are reported in red close to the corresponding block.}
    \label{fig_schema}
\end{figure*}

The flowchart of the calculations implemented for the time- and frequency-domain computation of the OIR increment $\Delta_{X^N_{-j};X_j}$ is depicted in Fig. \ref{fig_schema}. The procedure, which is described here making reference to the equations implemented in the various steps and to the corresponding codes of the OIR Matlab toolbox, starts with a vector process $Y$ organized in the blocks $\{X_1,\ldots,X_M\}$, and from a set of indices
identifying the group of block processes $X^N$ to analyze and the target process $X_j$.
The first steps are to identify the VAR model fitting the whole process $Y$ (\texttt{oir\_idVAR.m}, \texttt{oir\_mosVAR.m}, eq. \ref{VAR}) and to convert the VAR parameters into SS parameters (\texttt{oir\_ar2ss.m}, eq. \ref{ISS}). The SS parameters, together with the set of numbers indexing $X^N$ inside $Y$, are passed as inputs to the iteration computing the terms of the causal decomposition and frequency expansion of the MIR $I_{Z_1;Z_2}$ (\texttt{oir\_mir.m}). Such iteration performs, fixing $Z_1=X_j$ and varying $Z_2=X^N_{-mj}$ (\texttt{oir\_subindexes.m}), the extraction of the reduced model describing $Z=\{Z_1,Z_2\}$ and its conversion to SS form (\texttt{oir\_submodel.m}, eq. \ref{ISSZ}), followed by the computation of the frequency-domain coupling and causality measures (\texttt{oir\_fdGC.m}, eqs. \ref{ISSfreq}, \ref{f12}, \ref{fi_j}) and their integration to the time domain (eq. \ref{F12_f12}). All MIR measures are then combined in the frequency domain (eq. \ref{deltaOIRfreq}) and in the time domain (eq. \ref{deltaOIR}) to get the desired OIR increments (\texttt{oir\_deltaO.m}). These increments constitute the output of our algorithm, and can be easily exploited by the time-domain (eq. \ref{OIRrec}) and frequency-domain (eq. \ref{OIRrecfreq}) recursions to compute the OIR of any group of $N$ blocks extracted from the original vector process.

\section{Theoretical Examples} \label{sec_examples}

In this section, the framework for the computation of pairwise and higher-order interactions in the time and frequency domains is illustrated making use of theoretical examples of simulated multivariate VAR models for which the various measures are computed directly from the known model parameters. 
These simulations are exploited to show how our measures can be used:
(a) to highlight the emergence of patterns of interaction among groups of processes which cannot be traced from pairwise connections;
(b) to dissect pairwise and higher-order interactions into causal components which can be related to the topological structure of the underlying network;
(c) to ascribe interactions to specific oscillations confined within specific frequency bands;
(d) to evidence the presence of circuits dominated by synergy or redundancy, or even by simultaneous synergistic and redundant behaviors coexisting at different frequencies.
Detailed equations and parameter settings are provided for the two simulations in the supplemental material.

\subsection{Simulation 1} \label{sec_simu1}
\begin{figure} [t]
    \centering
    \includegraphics[scale=0.83]{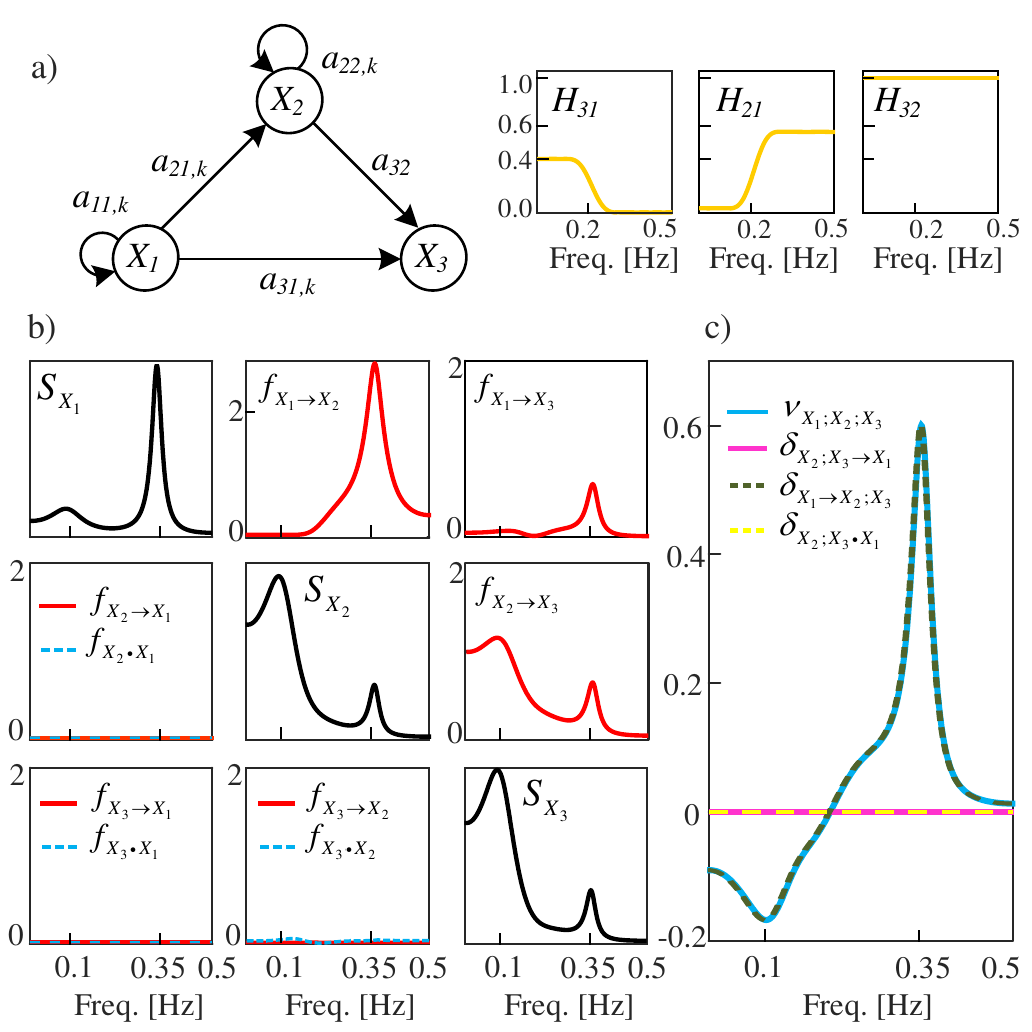}
    \caption{Theoretical simulation of cardiovascular interactions. (a) Connectivity structure of the simulated VAR process (left) and of its spectral transfer functions (right); (b) power spectral density of the three processes (diagonal) and components of the causal decomposition of the spectral coupling between each pair of processes (off-diagonal); (c) spectral profiles of the O-information rate of the three processes and of the components of its causal decomposition.}
    \label{fig_simu1}
\end{figure}

The first simulation reproduces the trivariate system proposed in \cite{faes2021information}, adapted to generate realistic cardiovascular and respiratory dynamics. The activity of this system is mapped by a trivariate VAR process defined as in (\ref{VAR}) fed by independent Gaussian innovations, for which the parameters are set as illustrated in Fig. \ref{fig_simu1}a.
The vector process is studied keeping the three scalar processes separate ($M=Q=3, X=Y$), and assuming sampling  frequency $f_s=1$ (spectral functions are described completely in the frequency range $0-0.5$ Hz).
The coefficient matrix $\mathbf{A}$ is designed to mimic the dynamics of respiration ($X_1$), arterial pressure ($X_2$) and heart period ($X_3$) variability, generating self-dependencies for the processes $X_1$ and $X_2$ through the coefficients $a_{11,k}$ and $a_{22,k}$, and imposing causal effects along the directions $X_1 \rightarrow X_2$, $X_1 \rightarrow X_3$ and  $X_2 \rightarrow X_3$ through the coefficients $a_{21,k}$, $a_{31,k}$ and  $a_{32}$.
Self-dependencies are set to induce oscillations in the respiratory band ($\sim 0.35$ Hz) for $X_1$ and in the low-frequency band ($\sim 0.1$ Hz) for $X_1$ and particularly for $X_2$, while causal effects are set to realize a high-pass filter from $X_1$ to $X_2$, a low-pass filter from $X_1$ to $X_3$ and an all-pass configuration from $X_2$ to $X_3$ (spectral transfer functions are shown in Fig. \ref{fig_simu1}a, right); low- and high-pass filtering are achieved through FIR filters of order 20 with cut-off frequency of 0.2 Hz.

The application of our framework to the VAR parameters describing the simulated process leads to the spectral functions depicted in Fig. \ref{fig_simu1}b,c. The PSD profiles (Fig. \ref{fig_simu1}b, diagonal plots) highlight oscillations at $\sim 0.1$ Hz and $\sim 0.35$ Hz for the three processes.
The causal coupling between pairs of processes (Fig. \ref{fig_simu1}b, off-diagonal plots) evidences the presence of information flows originating from the first process (nonzero profiles of $f_{X_1 \rightarrow X_2}$,$f_{X_1 \rightarrow X_3}$ and $f_{X_2 \rightarrow X_3}$) and the absence of information flowing back towards it ($f_{X_3 \rightarrow X_2}=f_{X_2 \rightarrow X_1}=f_{X_3 \rightarrow X_1}=0$ at each frequency). Note that, given the unidirectional coupling and the absence of instantaneous interactions, in virtue of (\ref{MIRexpZfreq}) the three nonzero causal coupling measures are equivalent to the spectral measures of total coupling $f_{X_1;X_2}$, $f_{X_1;X_3}$ and $f_{X_2;X_3}$ (red curves in Fig. \ref{fig_simu1}b); whole-band integration of such measures by (\ref{F12_f12}) leads to the MIR quantifying the total information shared between pairs of processes, whose values result $I_{X_1;X_2}=T_{X_1 \rightarrow X_2}=0.28$ nats, $I_{X_1;X_3}=T_{X_1 \rightarrow X_3}=0.05$ nats and $I_{X_2;X_3}=T_{X_2 \rightarrow X_3}=0.24$ nats.
Then, computation of the MIR between one process and the remaining two leads to obtain the OIR via (\ref{deltaOIR}), which for this simulation is $\Omega_{X_1;X_2;X_3}=0.019$ nats, denoting a small redundant interaction among the three processes. Importantly, the spectral expansion (Fig. \ref{fig_simu1}c) reveals that this small OIR value is the balance between a synergistic interaction at low frequencies ($\Omega_{X_1;X_2;X_3}=-0.15$ nats in the band $0.04-0.12$ Hz) and a redundant interaction at higher frequencies ($\Omega_{X_1;X_2;X_3}=+0.33$ nats in the band $0.31-0.39$ Hz). We also highlight that the causal decomposition of the OIR $\nu_{X_1;X_2;X_3}=\delta_{X_1;X_2,X_3}$ reveals the unidirectional nature of the OIR increment (i.e., $\delta_{X_1;X_2,X_3}=\delta_{X_1 \rightarrow X_2,X_3}$ and $\delta_{X_2,X_3 \rightarrow X_1}=\delta_{X_1 \boldsymbol{\bigcdot} X_2,X_3}=0$).
The opposite OIR values observed in the two frequency bands can be explained by the simulation design (see Fig. \ref{fig_simu1}a): synergy and redundancy arise respectively because the flow of information from $X_1$ to $X_3$ is entirely mediated by $X_2$ at the respiratory frequency (the path $X_1 \rightarrow X_3$ is blocked by $H_{31}$ at $\sim 0.35$ Hz), and because such flow occurs via the independent paths $X_1 \rightarrow X_3$ and $X_2 \rightarrow X_3$ at lower frequencies (the path $X_1 \rightarrow X_2$ is blocked by $H_{21}$ at $\sim 0.1$ Hz).

\subsection{Simulation 2} \label{sec_simu2}
\begin{figure*} [t]
    \centering
    \includegraphics[scale=0.87]{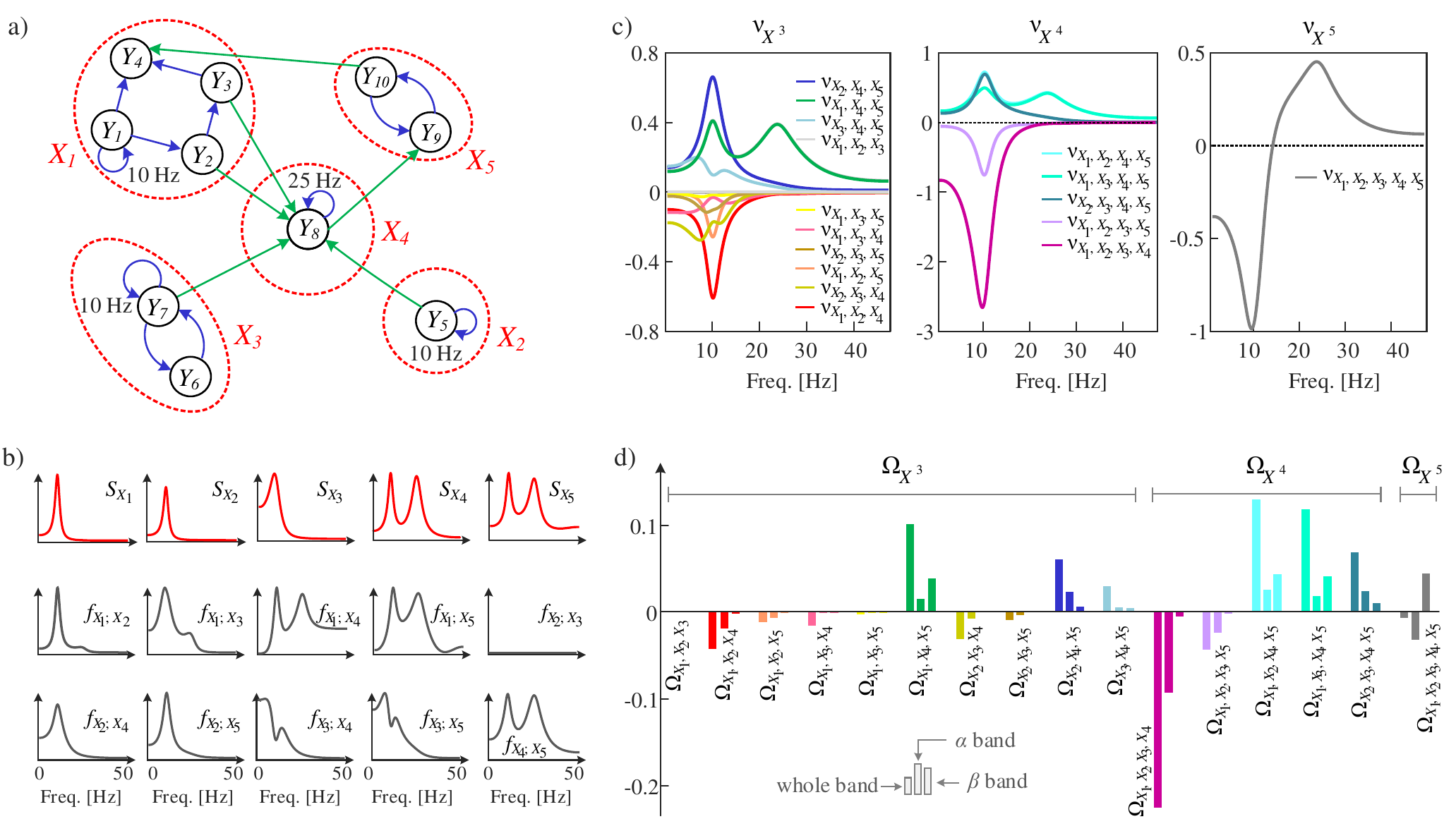}
    \caption{Theoretical simulation of neurophysiological interactions. (a) Connectivity structure of the simulated VAR process, featuring 10 scalar processes grouped in 5 blocks; (b) power spectral densities (red) and spectral coupling functions (gray) between each pair of block processes; (c) spectral profiles of the O-information rate computed for multiplets of three (left), four (middle) and five (right) block processes; (d) time-domain O-information rate obtained integrating the spectral measure relevant to each multiplet over the whole frequency axis (left bars), inside the $\alpha$ band (8-12 Hz, middle bars), or inside the $\beta$ band (18-30 Hz, right bars).}
    \label{fig_simu2}
\end{figure*}

The second simulation illustrates  the possibility offered by our framework to quantify higher-order spectral interactions among multiple blocks of processes whose dynamics resemble those of neurophyiological signals. The simulation extends previous simulations of VAR processes \cite{faes2013measuring,antonacci2021measuring} to the analysis of $Q=10$ processes organized in $M=5$ blocks, with connectivity structure organized as in Fig. \ref{fig_simu2}a.
The network is designed to simulate three autonomous vector processes $X_1$, $X_2$ and $X_3$ which generate, through their own subnetwork interactions, a stochastic oscillation resembling the brain $\alpha$ rhythm ($\sim 10$ Hz) which is transmitted to the central node $X_4$; such node is a sink for the $\alpha$ waves but also acts as a source of oscillatory activity in the $\beta$ band ($\sim 25$ Hz), which is transmitted back to $X_1$ through the passive block $X_5$.
The presence of the two simulated rhythms and their transmission through the network is documented by the power spectra $S_{X_i}$ and by the pairwise coupling measures $f_{X_i;X_j}$ reported respectively in red and gray in Fig. \ref{fig_simu2}b;
integration of the coupling measures leads to detect significant MIR values between each pair of processes except $X_2$ and $X_3$.

The analysis of higher-order interactions was performed computing the spectral OIR for all multiplets of order $N=3,4,5$ (Fig. \ref{fig_simu2}c) as well as the corresponding time-domain OIR values obtained integrating the spectral measures over all frequencies or within the $\alpha$ (8-12 Hz) or $\beta$ (18-30 Hz) bands (Fig. \ref{fig_simu2}d).
This analysis allows to evidence patterns of interaction which cannot be inferred from lower-order pairwise links.
In particular, the presence of independent sources sending information to a common target originates synergistic modes of interaction characterized by negative profiles of the OIR; this is the case for the multiplets including two or three of the source processes $X_1,X_2,X_3$ and one between $X_4$ and $X_5$ (e.g., $\nu_{X_1,X_2,X_4}$ and $\nu_{X_1,X_2,X_3,X_4}$, red and violet negative OIRs in Fig. \ref{fig_simu2}c).
On the contrary, chains of interactions including three or more block processes determine redundant modes of dependence characterized by positive OIR values; this occurs when one or two of the sources $X_1,X_2,X_3$ and both the driven processes $X_4$ and $X_5$ are included in the analyzed multiplet (e.g., $\nu_{X_1,X_4,X_5}$ and $\nu_{X_1,X_2,X_4,X_5}$, green and cyan positive OIRs in Fig. \ref{fig_simu2}c).
We note also that the OIR is uniformly null for the triplet with independent processes $\{X_1,X_2,X_3\}$ (gray line in Fig. \ref{fig_simu2}c, left panel).
The computation of the time-domain OIR puts in evidence the purely synergistic or redundant nature of the interactions occurring within the multiplets of order 3 and 4, as documented in Fig. \ref{fig_simu2}d by the clearly negative or positive values of the OIRs. Interestingly, the integration within a specific frequency band ($\alpha$ or $\beta$) leads to infer which is the rhythm mostly associated with the interactions, which in this simulation occur dominantly in the $\alpha$ band for the synergistic modes with negative OIR, and in both bands with prevalence of $\beta$ for the redundant modes with positive OIR.

The analysis of the highest-order multiplet incorporating all processes puts clearly in evidence that synergy and redundancy are related to the simulated $\alpha$ and $\beta$ rhythms, respectively. Indeed, the spectral OIR $\nu_{X^5}$ displays a negative peak at $\sim 10$ Hz and a positive peak at $\sim 25$ Hz (Fig. \ref{fig_simu2}c, right panel), and the integration of this spectral function within the $\alpha$ and $\beta$ bands evidences clearly negative and positive values (grey bars at the right of Fig. \ref{fig_simu2}d). This mode is an example of how the coexistence of synergy and redundancy at different frequencies may mask their time domain detection, as in this case the whole-band integration of the spectral OIR gives small negative values which could be difficult to assess in practice.

\section{Application to Physiological Networks} \label{sec_applications}
This section reports the application of the framework for the analysis of multivariate interactions in the time and frequency domain to two different physiological networks, i.e. cardiovascular and respiratory interactions during paced breathing, and neural interactions from ECoG signals in the anesthetized macaque monkey.
Full details about the analyzed datasets and complete results are provided in the supplemental material.

\subsection{Cardiovascular and respiratory interactions during paced breathing}
The analyzed dataset refers to beat-to-beat variability series of respiration (RESP, process $X_1$), systolic arterial pressure (SAP, process $X_2$) and heart period (HP, process$X_3$), synchronously measured in a group of 18 young healthy subjects monitored in the resting supine position during an experimental protocol consisting of four phases: spontaneous breathing (SB) and controlled breathing at 10, 15, and 20 breaths/minute (CB10, CB15, CB20) \cite{porta2011accounting}. The HP, SAP and RESP time series were extracted respectively from the electrocardiogram, noninvasive arterial blood pressure and nasal respiration flow as the sequences of the duration of the cardiac cycle (R-R interval), of the local maximum of the blood pressure signal within each detected cardiac cycle, and of the value of the respiration signal sampled at the onset of each cardiac cycle. This measurement convention implies that instantaneous influences can be described as causal effects from RESP to SAP and HP and from SAP to HP (directions $X_1 \rightarrow X_2$, $X_1 \rightarrow X_3$, $X_2 \rightarrow X_3$) \cite{faes2013framework}.

The analysis was performed on stationary segments of the time series including 256 heartbeats, selected by visual inspection for each subject and experimental condition  \cite{porta2011accounting}. The pre-processing consisted on detrending and mean removal for each time series.
The VAR model fitting the three series was identified through the ordinary least squares method, selecting the order $p$ in the range 3-14 by means of the Akaike Information Criterion \cite{faes2012measuring}.
The analysis was focused on decomposing the OIR of the three processes in OIR increments obtained when the HP process is added to the bivariate process $\{$RESP,SAP$\}$. Specifically, starting from the estimated VAR parameters, we computed $\delta_{X_1,X_2 \rightarrow X_3}(f)$, $\delta_{X_3 \rightarrow X_1,X_2}(f)$ and $\delta_{X_1,X_2\boldsymbol{\bigcdot}X_3}(f)$ from the terms of the spectral decomposition (10), then deriving $\nu_{X_1,X_2,X_3}(f)=\delta_{X_1,X_2;X_3}(f)$ via (13,14). From these spectral measures, time-domain measures were obtained through integration over the whole frequency axis or within the low frequency range (LF, 0.04-0.12 Hz) and the high frequency range (HF, $\pm 0.04$ Hz around the respiratory frequency $f_{RESP}$). 
Given the possibility to ascribe instantaneous effects to specific causal directions (see above), the analysis is performed summing the information shared instantaneously between $\{$RESP,SAP$\}$ and HP to the information transferred from $\{$RESP,SAP$\}$ to HP, i.e. computing the spectral and time domain measures
$\delta_{X_1,X_2 \overset{\text{\normalsize.}}\rightarrow X_3}(f)=\delta_{X_1,X_2 \rightarrow X_3}(f)+\delta_{X_1,X_2\boldsymbol{\bigcdot}X_3}(f)$ and 
$\Delta_{X_1,X_2 \overset{\text{\normalsize.}}\rightarrow X_3}=\Delta_{X_1,X_2 \rightarrow X_3}+\Delta_{X_1,X_2\boldsymbol{\bigcdot}X_3}$.

\begin{figure} 
    \centering
    \includegraphics[scale=0.8]{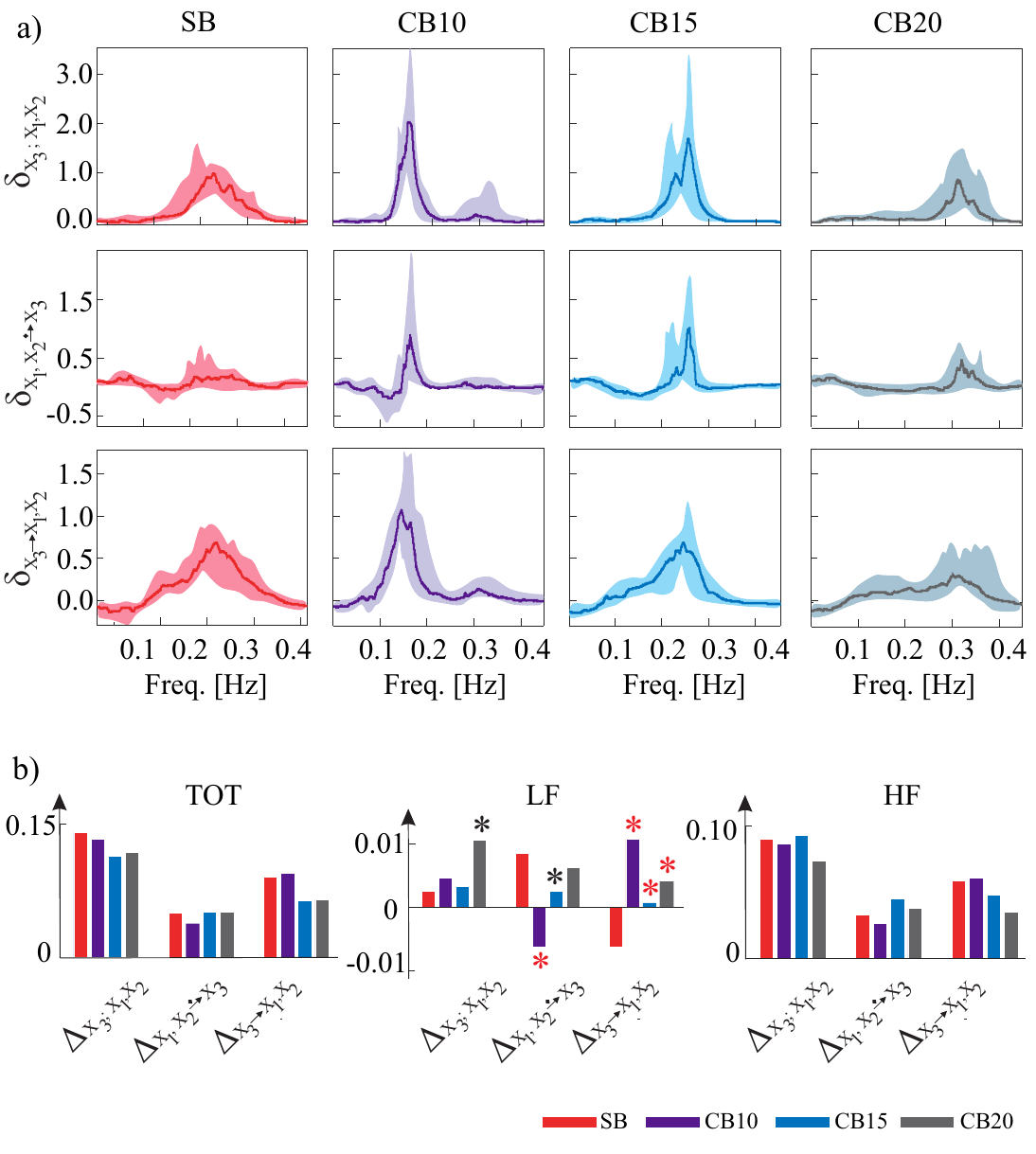}
    \caption{OIR decomposition of cardiovascular interactions during controlled breathing (CB). (a) Average spectral profiles across subjects (line: median; shades: 1st-3rd quartiles) of the OIR increment obtained with the addition of HP to $\{$SAP,RESP$\}$ (upper panels) and of its decomposition in causal terms (middle and lower panels) computed during spontaneous breathing (SB) and CB at 10, 15 and 20 breaths/min. (b) Time-domain values of the mean OIR increments obtained integrating the spectral measures over the whole frequency axis (TOT), in the range 0.04-0.12 Hz (LF) or in the range $f_{RESP}\pm0.04$ Hz (HF); asterisks denote statistically significant difference between the CB condition compared with SB (Wilcoxon signed-rank test: black, uncorrected; red, Bonferroni-Holm correction for multiple comparisons).}
    \label{fig_cardio}
\end{figure}

The results of OIR computation and decomposition are reported in Fig. \ref{fig_cardio}, showing the grand average of the frequency-domain measures as well as the whole-band, LF and HF time-domain average measures. Spectral analysis was performed assuming the series as uniformly sampled with sampling frequency equal to the inverse of the mean HP. The spectral OIR and most of the terms of its decomposition exhibit prominent peaks, which are well-defined at the frequency of the paced breathing during the CB conditions and are less narrow-banded during SB (Fig. \ref{fig_cardio}a). This behavior reflects the fact that paced breathing regularizes the RESP signal around the imposed rhythm and enforces synchronous oscillations at the same frequency in the HP and SAP time series, determining increased spectral content and spectral coupling in the HF band \cite{porta2011accounting}. The positive values of the time-domain OIR (Fig. \ref{fig_cardio}b, left) document that this synchronized interaction is dominantly redundant, confirming previous findings \cite{faes2017information}. Looking at the spectral profiles of Fig. \ref{fig_cardio}a, the peak values of the OIR show a tendency to increase while moving from SB to CB10, and to decrease progressively during CB15 and CB20; these trends confirm from the perspective of HOIs results obtained on the same data using information-theoretic measures of cardiorespiratory coupling \cite{porta2000information}.
The dominance of redundancy in the HF band of the spectrum (Fig. \ref{fig_cardio}b, right) suggests that the main underlying physiological mechanism is the mechanical influence of RESP on SAP variability, transmitted to HP through the baroreflex feedback \cite{krohova2019multiscale}; the OIR component directed from HP to $\{$SAP,RESP$\}$, which tends to be less redundant at increasing the frequency of paced breathing, is of more difficult interpretation and is likely dominated by the mechanical feedforward effects from HP to SAP \cite{javorka2017causal}. 
The dominance of redundant mechanisms around the respiratory frequency impacts substantially the whole-band time-domain OIR, which show comparable values across the analyzed conditions (Fig. \ref{fig_cardio}b, left).
On the other hand, the measures integrated within the LF band vary significantly moving from spontaneous to paced breathing (Fig. \ref{fig_cardio}b, middle): the information transfer from $\{$SAP,RESP$\}$ to HP becomes mostly synergistic during CB10, and during CB15 and CB20 returns progressively to the redundant values observed at SB; the information transfer along the direction HP$\rightarrow\{$SAP,RESP$\}$ is prevalently synergistic at rest and shifts to redundant values during CB. The shift to synergy observed at CB10 for $\Delta_{X_1,X_2\rightarrow X_3}$ suggests that, when the respiratory activity slows down and tends to overlap with the Mayer waves typically observed in SAP and HP \cite{julien2006enigma}, the baroreflex (SAP$\rightarrow$HP) and respiratory sinus arrhythmia (RESP$\rightarrow$HP) mechanisms operate independently in determining the variability of heart rate.

\subsection{Neural interactions from ECoG signals in the anesthetized macaque monkey}
\begin{figure} [h!]
    \centering
    \includegraphics[scale=0.87]{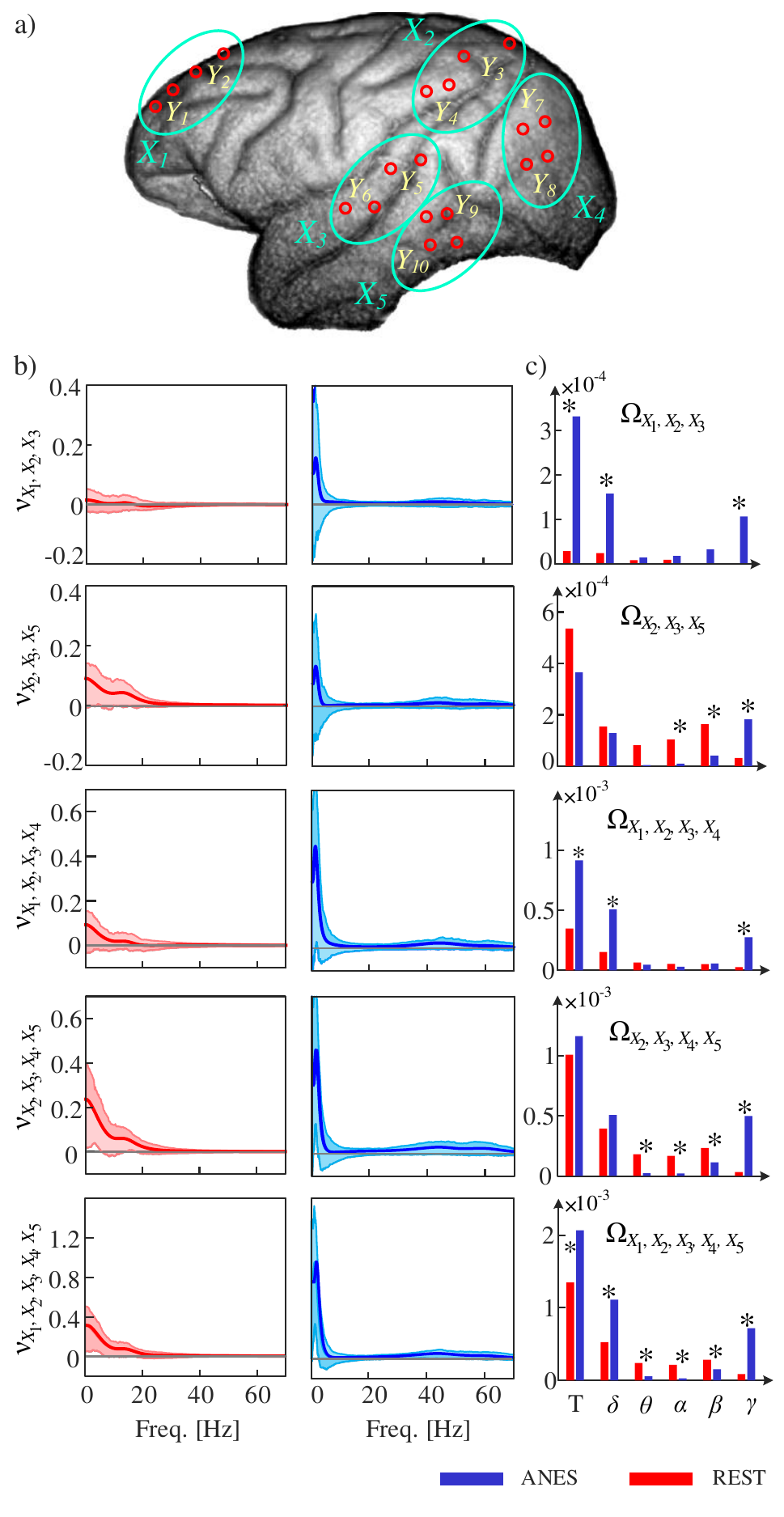}
    \caption{OIR analysis of neurophysiological interactions in the anesthetized monkey. (a) ECoG electrode montage highlighting the positions of the selected electrodes acquiring the bipolar signals $Y_1,\ldots, Y_{10}$ grouped in the blocks $X_1, \ldots, X_5$ covering five regions of the left hemisphere. (b) Average spectral profiles across trials (line: median; shades: 1st-3rd quartiles) of the OIR computed for five representative multiplets during relaxation (REST) and  anesthesia (ANES). (c) Time-domain values of the mean OIR obtained by integrating the spectral measures over the whole frequency axis (T) or within the $\delta$, $\theta$, $\alpha$, $\beta$ and $\gamma$ bands; asterisks  denote statistically significant difference between REST and ANES (Wilcoxon signed-rank test with Bonferroni correction for multiple comparisons ).}
    \label{fig_ECoG}
\end{figure}

The second practical application refers to monkey electrocorticographic (ECoG) signals downloaded from the public server neurotycho.org. The analyzed dataset was recorded with a sampling frequency of 1000 Hz in one macaque monkey using 128  electrodes, placed in pairs with an inter-electrode distance of 5 mm to cover the frontal, parietal, temporal and occipital lobes of the left hemisphere \cite{yanagawa2013large}.
Specifically, we considered two five-minutes recording sessions during which the blindfolded monkey was seated in a primate chair with tied hands, first in a resting state (REST) and then after injection of a sedative inducing anesthesia (ANES).
From the 128 electrodes, a subset of 20 was selected as depicted in Figure \ref{fig_ECoG}a to cover, considering ten bipolar ECoG signals obtained taking the differential activity between close electrodes, the following five brain regions of the default mode network, i.e. the pre-frontal cortex ($X_1=[Y_1,Y_2]$), parietal cortex ($X_2=[Y_3,Y_4]$), temporal cortex ($X_3=[Y_5,Y_6]$), low visual cortex ($X_4=[Y_7,Y_8]$), and high visual cortex ($X_5=[Y_9,Y_{10}]$).
The ten bipolar signals were band-pass filtered between 0.5 and 200 Hz, 
downsampled to $f_s=250$ Hz, epoched to extract $\sim 160$ trials lasting 2 sec for each condition, and finally normalized to zero mean and unit variance within each trial. Then, a VAR model was fitted on the $Q=10$ signals of each trial using least squares identification and setting the model order according to the Bayesian Information Criterion (BIC) \cite{faes2012measuring}. From the VAR parameters, the analysis of high-order interactions was performed for the $M=5$ blocks computing the spectral OIR for all multiplets of order $N=3,4,5$. Time-domain OIR values ($\Omega$) were then obtained integrating the spectral measures $\nu(f)$ within the $\delta$ (0.2-3 Hz), $\theta$ (4-7 Hz), $\alpha$ (8-12 Hz), $\beta$ (12-30 Hz) and $\gamma$ (31-70 Hz) frequency bands, as well as cumulatively between 0 and 70 Hz.

The results of OIR computation are reported in Fig. \ref{fig_ECoG}b, showing the grand average of the spectral OIR for five multiplets selected as the most representative of the analyzed interactions, together with the time-domain OIR obtained through whole-band  and band-specific integration.
The positive values of the OIR functions and of the integrated measures, observed for all multiplets in both conditions and increasing with the order of the multiplet, indicate that the analyzed system is dominated by redundancy.
Moreover, the redundancy level is modulated by the experimental condition to an extent that depends on the analyzed multiplet and spectral band. Indeed, considering the multiplets of order 3 and 4 which involve the prefrontal cortex $X_1$ (1st and 3rd row of panels in Fig. \ref{fig_ECoG}b), a significant increase of the OIR is observed while moving from REST to ANES; such increase is driven by the rise of a peak in the OIR at $\sim 2$ Hz ($\delta$ band) together with an increased contribution within the $\gamma$ band. On the other hand, the multiplets formed by signals from the parietal, temporal and visual cortices  (2nd and 4th row of panels in Fig. \ref{fig_ECoG}b) display a drop of redundancy in the $\alpha$ and $\beta$ bands during ANES compared to REST. These two opposite behaviors are summarized by the OIR encompassing all five regions (5th row of panels in Fig. \ref{fig_ECoG}b), which during ANES displays significantly higher levels of redundancy in the $\delta$ and $\gamma$ bands (and in the whole band), and significantly lower redundancy in the $\theta$, $\alpha$, and $\beta$ bands. 


Our results indicate that the activity relevant to the $\alpha$ and $\beta$ rhythms observed  during the relaxed awake state disappears during anesthesia, leaving place to dominant interactions within the $\delta$ and $\gamma$ bands. The redundancy observed at REST for the $\alpha$ waves is significant for the multiplets involving signals from the visual cortex, in agreement with the knowledge that these waves can be predominantly recorded from the occipital lobes during wakeful relaxation with closed eyes \cite{palva2007new}. On the other hand, the higher redundancy reported in the $\delta$ band can be related to the slow wave oscillations (0.1-4 Hz) typically observed under anesthesia \cite{chauvette2011properties}. Moreover, the fact that higher $\delta$ redundancy is observed only for multiplets including frontal cortex signals supports the knowledge that the slow oscillations  are a manifestation of a coupling between the anterior and posterior axes of the brain \cite{murphy2009source}. Anesthesia evokes also an increase of redundancy related to $\gamma$ oscillations, which are associated with different cognitive functions \cite{fries2009neuronal}.

Overall, these results agree with those in \cite{yanagawa2013large} and support the integration theory according to which the conscious state is generated by highly integrated neural interactions that disappear in the unconscious state \cite{baars2002conscious}.  
A recent study comparing resting wakefulness with propofol-induced anaesthesia in human fMRI data has shown how the anterior-posterior disconnection occurring during anesthesia is associated with a decrease of Integrated Information within the default mode network in the left hemisphere  \cite{luppi2020synergistic}. Importantly, the concepts of Integration Information and that of redundancy are interrelated, as explained in \cite{mediano2019beyond} where it is highlighted that a drop of Integrated Information corresponds to an increase of redundancy.
Thus, our results support the theory of an anterior-posterior disconnection during anesthesia, which in our case can be ascribed to the significant increase of the OIR documented when the frontal cortex is considered in the analyzed multiplet. 

\section{Conclusion}
This work opens the way to the combined information-theoretic and spectral evaluation of hierarchically-organized interactions in dynamic networks mapped by multivariate stochastic processes.
The proposed framework is highly flexible and scalable as it provides principled measures of both pairwise and higher-order interactions among scalar or vector processes, defined in both time and frequency domains in a way such that the two representations are connected in a straightforward way. Moreover, it allows to decompose symmetric measures into components reflecting Granger-causal and instantaneous influences, and to estimate them with high computational reliability within the framework of vector autoregressive and state space models.

The application of the new framework to biomedical time series illustrates its capability to capture the balance between redundancies and synergies among arbitrarily large groups of nodes of brain and physiological networks. Moreover, it highlights the importance of studying these features within specific frequency bands of biological interest to elicit interactions which may be otherwise hidden if investigated only in the time domain.
The generality of the information-theoretic grounds and of the parametric implementation of the proposed approach makes it suitable for the assessment of pairwise and higher-order interactions even beyond the domain of biomedical time series, to analyze virtually any type of dynamic network (e.g., electronic, climatologic, social, or financial) with node activity described by rhythmic processes.





\bibliography{OIRref}

\pagebreak
\widetext
\begin{center}
\textbf{\large Supplemental material:  Framework for the Time- and Frequency-Domain Assessment of High-Order Interactions in Brain andPhysiological Networks}
\end{center}

\section*{Framework to measure High-Order Interactions in multivariate processes}

\subsection{O-information rate}
\label{sec21}
Considering a generic stationary stochastic process $X_i$ 
composed by the random variables $X_{i,n}$ (where $n \in \mathbb{N}$ is the temporal index), the following definitions of entropy rate are equivalent under the assumption of stationarity \cite{cover1999elements}:
\begin{equation}
   H_{X_i}=\lim_{k \to \infty} \dfrac{1}{k}H(X_{i,n}^k)
   =H(X_{i,n}|X_{i,n}^-), \label{HR}
\end{equation}
where $H(X_{i,n}^k)$ is the entropy of $X^k_{i,n}=[X_{i,n-1} \cdots X_{i,n-k}]$ and $H(X_{i,n}|X_{i,n}^-)$ is the conditional entropy of $X_{i,n}$ given $X^-_{i,n}=\lim_{k \to \infty}X^k_{i,n}$. With our notation, $H(\bigcdot)$ denotes the entropy of a random variable, and $H_{(\bigcdot)}$ denotes the entropy rate of a random process.
As (\ref{HR}) holds also for groups of processes, the definition of mutual information (MI) \cite{cover1999elements} can be exploited to define the MI rate (MIR) between the processes $X_i$ and $X_j$ as \cite{duncan1970calculation}
\begin{equation}
   I_{X_i;X_j}=\lim_{k \to \infty} \dfrac{1}{k}I(X_{i,n}^k;X_{j,n}^k)=H_{X_i}+H_{X_j}-H_{X_i,X_j} \label{MIR}.
\end{equation}
The O-information rate (OIR) of a set of $N$ stochastic processes $X^N=\{X_1,...,X_N\}$ is defined elaborating the entropy rates of subsets of $X^N$ according to
\begin{equation}
   \Omega_{X^N}=(N-2)H_{X^N} + \sum_{i=1}^{N}[H_{X_i}-H_{X^N_{-i}}]
   \label{OIR},
\end{equation}
where $X^N_{-i}=X^N \backslash X_i$ is the subset of processes where $X_i$ is removed. The OIR defined in (\ref{OIR}) is a symmetric measure which generalizes to stochastic processes the O-information measure recently proposed to assess the "organization structure" of a group of random variables \cite{rosas2019quantifying}.
For a bivariate process ($N=2$), it is easy to show from (\ref{OIR}) that $\Omega_{X^2}=0$. For $N=3$ processes, the OIR is equivalent to the interaction information rate, i.e. $\Omega_{X^3}=I_{X_j;X_i}+I_{X_j;X_k}-I_{X_j;X_i,X_k}$ (with $i,j,k=\{1,2,3\}$), a measure which generalizes to stochastic processes the concept of interaction information \cite{mcgill1954multivariate}. Then, applying to information rates the extensions provided in \cite{rosas2019quantifying} for information quantities, one can derive the iterative definition of the OIR given in Eqs. 1 and 2 of the main paper (see also \cite{stramaglia2012expanding}), which can be used to derive, for any order $N \geq 3$, the OIR of order $N$ given any OIR of order $N-1$ and the corresponding OIR increment:
\begin{subequations}
\begin{align}
   \Omega_{X^N}&=\Omega_{X^N_{-j}}+\Delta_{X^N_{-j};X_j},\\
   \Delta_{X^N_{-j};X_j}&=(2-N)I_{X_j;X^N_{-j}} + \sum_{\substack{m=1 \\ m \neq j}}^{N}I_{X_j;X^N_{-mj}}.
   \label{DeltaOIR}
\end{align}
\label{OIRrec}
\end{subequations}


\subsection{Causal decomposition of the O-information rate}
\label{sec22}
Given two processes $X_i$ and $X_j$, the MIR (\ref{MIR}) can be formulated according to the expansion (see, e.g., \cite{chicharro2011spectral})
\begin{equation}
   I_{X_i;X_j}=T_{X_j \rightarrow X_i}+T_{X_i \rightarrow X_j}+I_{X_i \boldsymbol{\bigcdot} X_j}
   \label{MIRexp},
\end{equation}
where $T_{X_j \rightarrow X_i}=I(X_{i,n};X_{j,n}^-|X_{i,n}^-)$ and $T_{X_i \rightarrow X_j}=I(X_{j,n};X_{i,n}^-|X_{j,n}^-)$ are the transfer entropy from $X_j$ to $X_i$ and from $X_i$ to $X_j$, and $I_{X_i \bigcdot X_j}=I(X_{i,n};X_{j,n}|X_{i,n}^-,X_{j,n}^-)$ represents the instantaneous information shared between $X_i$ and $X_j$ (where $I(\cdot;\cdot|\cdot)$ denotes conditional MI for three random variables).
Since the decomposition (\ref{MIRexp}) is equally valid for groups of stochastic processes, it can be applied to any MIR term appearing in the OIR increment defined in \ref{DeltaOIR}, which can be therefore rewritten in the form of Eq. 4 of the main paper as
\begin{equation}
   \Delta_{X^N_{-j};X_j} = \Delta_{X^N_{-j} \rightarrow X_j} + \Delta_{X_j \rightarrow X^N_{-j}} + \Delta_{X^N_{-j} \bigcdot X_j}
   \label{deltaOIRexp},
\end{equation}
where the three terms
\begin{subequations}
\begin{align}
\Delta_{X^N_{-j} \rightarrow X_j} &= (2-N)T_{X^N_{-j} \rightarrow X_j} + \sum_{\substack{m=1 \\ m \neq j}}^{N}T_{X^N_{-mj} \rightarrow X_j} \label{deltaOIR1}\\
\Delta_{X_j \rightarrow X^N_{-j} } &= (2-N)T_{X_j \rightarrow X^N_{-j}} + \sum_{\substack{m=1 \\ m \neq j}}^{N}T_{X_j \rightarrow X^N_{-mj}} \label{deltaOIR2}\\
\Delta_{X^N_{-j} \bigcdot X_j} &= (2-N)I_{X^N_{-j} \bigcdot X_j} + \sum_{\substack{m=1 \\ m \neq j}}^{N}I_{X^N_{-mj} \bigcdot X_j} \label{deltaOIR3}
\end{align}
\label{deltaOIRterms}
\end{subequations}
quantify the informational character of the directed information transfer from $X^N_{-j}$ to $X_j$, of the directed information transfer from $X_j$ to $X^N_{-j}$, and of the  instantaneous information shared between $X_j$ and $X^N_{-j}$, respectively; the informational character of each term is redundant when the term is positive, and synergistic when the term is negative.

\subsection{Linear parametric formulation}
\label{sec23}
The OIR framework considers a network of $Q$ stationary stochastic processes $Y=\{Y_1, \ldots, Y_Q\}$, whose dynamics are represented by the linear parametric model given by Eq. 5 of the main paper:
\begin{equation}
   Y_n = \sum_{k=1}^{p}\mathbf{A}_k Y_{n-k} + U_n
   \label{VAR}.
\end{equation}
In our analyses, the original processes are grouped into $M$ blocks $X_1, \ldots, X_M$, from which the processes $Z_1=X_j$ and $Z_2=X^N_{-mj}$ of dimension $R_1$ and $R_2$ are selected (varying $m$ in the range $\{0,1,\ldots,M$, all processes appearing in \ref{DeltaOIR} can be considered). The linear parametric description of the vector process $Z=\{Z_1,Z_2\}$ is given by the VAR sub-model in Eq. 6 of the main paper,
\begin{equation}
   Z_n = \sum_{k=1}^{\infty}\mathbf{B}_k Z_{n-k} + W_n
   \label{VARZ},
\end{equation}
where $Z_n=[Z_{1,n}^\intercal Z_{2,n}^\intercal]^\intercal$ and $W_n=[W_{1,n}^\intercal W_{2,n}^\intercal]^\intercal$ are vectors of dimension $R=R_1+R_2$, and the $R \times R$  innovation covariance $\mathbf{\Sigma}_W=\mathbb{E}[W_n W_n^\intercal]$ is a block matrix having the $R_1 \times R_1$ matrix $\mathbf{\Sigma}_{W_{11}}=\mathbb{E}[W_{1,n} W_{1,n}^\intercal]$ and the $R_2 \times R_2$ matrix $\mathbf{\Sigma}_{W_{22}}=\mathbb{E}[W_{2,n} W_{2,n}^\intercal]$ as diagonal blocks.
The model (\ref{VARZ}) that provides a joint description of $Z=\{Z_1,Z_2\}$ is denoted as "full" model; then, the processes $Z_1$ and $Z_2$ can be described individually by the "restricted" models
\begin{equation}
   Z_{i,n} = \sum_{k=1}^{\infty}\mathbf{C}_{i,k} Z_{i,n-k} + V_{i,n}, i \in \{1,2\}
   \label{VARZrestricted},
\end{equation}
for which $\mathbf{C}_{i,k}$ is an $R_i \times R_i$ coefficient matrix and $V_i$ is an $R_i$-dimensional zero-mean white noise process with covariance matrix $\mathbf{\Sigma}_{V_i}=\mathbb{E}[V_{i,n} V_{i,n}^\intercal]$ of dimension $R_i \times R_i$. The covariance matrices of the full and restricted models (\ref{VARZ}) and (\ref{VARZrestricted}) are exploited to define measures of global, causal and  instantaneous interdependence between $Z_1$ and $Z_2$ as 
\begin{subequations}
\begin{align}
I_{Z_1;Z_2} &= \frac{1}{2} \log \frac{|\mathbf{\Sigma}_{V_1}| |\mathbf{\Sigma}_{V_2}|}{|\mathbf{\Sigma}_{W}|}, \label{F12gauss}\\
T_{Z_1 \rightarrow Z_2} &= \frac{1}{2} \log \frac{|\mathbf{\Sigma}_{V_2}|}{|\mathbf{\Sigma}_{W_{22}}|}  ,  T_{Z_2 \rightarrow Z_1} = \frac{1}{2} \log \frac{|\mathbf{\Sigma}_{V_1}|}{|\mathbf{\Sigma}_{W_{11}}|}, \label{F1_2gauss}\\
I_{Z_1 \boldsymbol{\bigcdot} Z_2} &= \frac{1}{2} \log \frac{|\mathbf{\Sigma}_{W_{11}}| |\mathbf{\Sigma}_{W_{22}}|}{|\mathbf{\Sigma}_{W}|}, \label{F1o2gauss}
\end{align}
\label{MIRexpGaussian}
\end{subequations}
which satisfy Eq. 3 of the main paper when the observed processes have a joint Gaussian distribution \cite{barnett2009granger}:
$I_{Z_1;Z_2}=T_{Z_1 \rightarrow Z_2}+T_{Z_2 \rightarrow Z_1}+I_{Z_1 \boldsymbol{\bigcdot} Z_2}$.
Note that this decomposition is the same as that defined here in (\ref{MIRexp}), generalized to the generic vector processes $Z_1$ and $Z_2$. The measures (\ref{MIRexpGaussian}), defined without the multiplicative factor $\frac{1}{2}$, were originally proposed by \cite{geweke1982measurement} in the framework of linear prediction. Here, exploiting the equivalence (up to a factor two) of linear prediction and information-theoretic measures valid for Gaussian processes \cite{barnett2009granger,barrett2010multivariate}, we have provided their direct formulation as measures of MIR (\ref{F12gauss}), transfer entropy (\ref{F1_2gauss}), and instantaneous information transfer (\ref{F1o2gauss}).

The derivations above show that, for Gaussian systems, the decomposition of MIR and the consequent computation of any OIR measure can be performed using nested linear regression models.
For these models, all the partial covariance measures (i.e., the determinants of the innovation covariance matrices appearing in (\ref{MIRexpGaussian})) which are needed for the computation of MIR and OIR can be obtained from the parameters of the original VAR process (\ref{VAR}) after expressing this process as a state space (SS) process.
To do this, we define the $pQ$-dimensional state process $S_n=[Y_{n-1}^\intercal \cdots Y_{n-p}^\intercal]^\intercal$ that, together with $Y_n$, obeys the equations of the SS model defined in Eq. 15 of the main paper,
\begin{subequations} \label{ISS}
\begin{align}
		S_{n+1} &= \mathbf{A} S_{n} + \mathbf{K} U_{n}, \label{ISSstate} \\
		Y_n &= \mathbf{C} S_{n} + U_{n}. \label{ISSobs}
\end{align}
\end{subequations}
The parameters of the SS model (\ref{ISS}) are ($\mathbf{A},\mathbf{C},\mathbf{K},\mathbf{V}$), of dimension $(pQ \times pQ, Q\times pQ, pQ \times Q, Q \times Q)$, where 
\[\mathbf{A}
=
\begin{bmatrix}
    \mathbf{A}_1&\cdots&\mathbf{A}_{p-1}&\mathbf{A}_p \\
		\mathbf{I}_Q&\cdots&\mathbf{0}_Q    &\mathbf{0}_Q \\
		\vdots      &      &\vdots          &\vdots       \\
		\mathbf{0}_Q&\cdots&\mathbf{I}_Q    &\mathbf{0}_Q\\
\end{bmatrix},
\]
$\mathbf{C}=[\mathbf{A}_1\cdots\mathbf{A}_p]$, $\mathbf{K}=[\mathbf{I}_Q \mathbf{0}_{Q\times Q(p-1)}]^\intercal$, and $\mathbf{V}=\mathbb{E}[U_n U_n^\intercal]=\mathbf{\Sigma}_U$.
Then, to represent the $R$-dimensional process $Z=\{Z_1,Z_2\}$
we replace (\ref{VARZ}) with a reduced SS model which has the same state equation of the overall model and a reduced observation equation, i.e. the model
\begin{subequations} \label{SSZ}
\begin{align}
		S_{n+1} &= \mathbf{A} S_{n} + \mathbf{K} U_{n}, \label{ISSZstate} \\
		Z_n &= \mathbf{C}^{(\mathbf{r},:)} S_{n} + W_{n}. \label{ISSZobs}
\end{align}
\end{subequations}
The parameters of the model (\ref{SSZ}) are ($\mathbf{A},\mathbf{C}^{(\mathbf{r},:)},\mathbf{K}\mathbf{V}\mathbf{K}^\intercal,\mathbf{V}^{(\mathbf{r},\mathbf{r})},\mathbf{K}\mathbf{V}^{(:,\mathbf{r})}$), where the superscripts $^{(\mathbf{r},:)}$, $^{(:,\mathbf{r})}$, and $^{(\mathbf{r},\mathbf{r})}$ denote selection of the rows and/or columns with indices $\mathbf{r}$ in a matrix.
Since the model  (\ref{SSZ}) has a different form than the original SS model, to exploit this model for the Granger-causal analysis of $Z$ it is necessary to lead its form back to that of (\ref{ISS}) \cite{barnett2015granger}, which gives Eq. 16 of the main paper:
\begin{subequations} \label{ISSZ}
\begin{align}
		S_{n+1} &= \Tilde{\mathbf{A}} S_{n} + \Tilde{\mathbf{K}} W_{n}, \label{ISSZstate} \\
		Z_n &= \Tilde{\mathbf{C}} S_{n} + W_{n}. \label{ISSZobs}
\end{align}
\end{subequations}
The parameters of the model (\ref{ISSZ}) are ($\Tilde{\mathbf{A}},\Tilde{\mathbf{C}},\Tilde{\mathbf{K}},\Tilde{\mathbf{V}}$), of dimension $(pQ \times pQ, R\times pQ, pQ \times R, R \times R)$; two of them can be retrieved directly from the original SS parameters, i.e. $\Tilde{\mathbf{A}}=\mathbf{A}$ and $\Tilde{\mathbf{C}}=\mathbf{C}^{(\mathbf{r},:)}$, while the gain $\Tilde{\mathbf{K}}$ and the reduced innovation covariance $\Tilde{\mathbf{V}}=\mathbb{E}[W_n W_n^\intercal]=\mathbf{\Sigma}_W$ can be obtained from the parameters in (\ref{ISS}) and (\ref{SSZ}) by solving a discrete algebraic Riccati equation (DARE) (see, e.g., refs. \cite{barnett2009granger,faes2017multiscale} for detailed derivations).

The computation of Granger-causal measures requires to formulate two additional reduced models, i.e. those describing the blocks $Z_1$ and $Z_2$ which compose the reduced process $Z$. These models, which are defined and identified as described above using the subprocesses $Z_1$ or $Z_2$ in place of $Z$ in (\ref{SSZ}) and (\ref{ISSZ}), constitute the SS representation of the restricted VAR models formulated in (\ref{VARZrestricted}). Their identification solving the DARE equation returns, among the other parameters, the covariance matrices of the innovations $V_1$ and $V_2$, which can be used in (\ref{MIRexpGaussian}) together with the covariance of the innovations $W$ to derive the terms of the causal decomposition of the MIR between $Z_1$ and $Z_2$.
We note that this procedure is alternative to the one working in the frequency domain described in the main paper, which is presented in extended form in the next section of this Supplemental Material. The advantage of the procedure described here is that it does not need to translate the models in the frequency domain and then to provide full spectral integration to retrieve the time-domain measures, while the disadvantage is that it requires the formulation of the two additional restricted models to derive all the partial variances appearing in (\ref{MIRexpGaussian}).

\subsection{Frequency domain expansion}
\label{sec24}
Starting from the subset  $Z=\{Z_1,Z_2\}$ of the observed multivariate process, the joint parametric description (\ref{VARZ}) can be expressed in the frequency domain taking the FT to obtain Eq. 7 of the main paper:
\begin{equation}
   Z(\omega) = [\mathbf{I}_R - \sum_{k=1}^{\infty}\mathbf{B}_k e^{-\mathbf{j} \omega k}]^{-1} W(\omega)=\mathbf{H}(\omega)W(\omega)
   \label{VARfreq}.
\end{equation}
Eq. (\ref{VARfreq}) allows to obtain the transfer function matrix $\mathbf{H}(\omega)$ starting from the parameters $\mathbf{B}_k$ of the reduced VAR model (\ref{VARZ}). However, since the identification of this infinite-order model is impractical, we replace it with the identification of the reduced model (\ref{ISSZ}); the frequency domain analysis of this model is performed taking the FT of (\ref{ISSZstate}), which reads
\begin{equation}
   S(\omega) = \Tilde{\mathbf{A}} S(\omega) + \Tilde{\mathbf{K}} W(\omega) e^{-\mathbf{j} \omega},
   \label{FTSeq}
\end{equation}
from which it is easy to derive the PSD of the state process, $S(\omega)$,
and to substitute it in the FT of (\ref{ISSZobs}) to obtain Eq. (17) of the main paper:
\begin{equation}
   Z(\omega) = \big(\mathbf{I}_R + \Tilde{\mathbf{C}}[\mathbf{I}_{pQ}-\Tilde{\mathbf{A}}e^{-\mathbf{j} \omega}]^{-1}\Tilde{\mathbf{K}}e^{-\mathbf{j} \omega}\big)W(\omega)=\mathbf{H}(\omega)W(\omega)
   \label{ISSfreq}.
\end{equation}

Then, given transfer function matrix $\mathbf{H}(\omega)$, the spectral factorization theorem allows to derive the PSD of the process using also the innovation covariance matrix \cite{anderson1982identifiability}:
\begin{equation}
   \mathbf{S}_Z(\omega)=\mathbf{H}(\omega)\mathbf{\Sigma}_{W}\mathbf{H}^*(\omega)
   \label{PSDspectdec}.
\end{equation}
The matrix $\mathbf{S}_Z(\omega)$ can be factorized in blocks to make explicit the power spectral densities of $Z_1$ and $Z_2$, $\mathbf{S}_{Z_1}(\omega)$ and $\mathbf{S}_{Z_2}(\omega)$, as diagonal blocks, and the cross-spectral densities between $Z_1$ and $Z_2$, $\mathbf{S}_{Z_1Z_2}(\omega)$ and $\mathbf{S}_{Z_2Z_1}(\omega)=\mathbf{S}_{Z_1Z_2}^*(\omega)$, as off-diagonal blocks. From this factorization, a logarithmic spectral measure of the interdependence between $Z_1$ and $Z_2$ is defined by Eq. 8 of the main paper \cite{geweke1982measurement}:
\begin{equation}
    f_{Z_1;Z_2}(\omega)=\log \frac{|\mathbf{S}_{Z_1}(\omega)||\mathbf{S}_{Z_2}(\omega)|}{|\mathbf{S}_{Z}(\omega)|};
   \label{f12}
\end{equation}
this measure quantifies the total (symmetric) coupling between the two block processes, and is related to the so-called block coherence \cite{nedungadi2011block}, which extends to vector processes the standard spectral coherence function \cite{marple1989digital}.
Moreover, after using (\ref{PSDspectdec}) to expand the PSD of $Z_i$, $i=1,2$, as
\begin{equation}
   \mathbf{S}_{Z_i}(\omega)=\mathbf{H}_{ii}(\omega)\mathbf{\Sigma}_{W_i}\mathbf{H}_{ii}^*(\omega) +
   \mathbf{H}_{ij}(\omega)\mathbf{\Sigma}_{W_j}\mathbf{H}_{ij}^*(\omega) +
   \mathbf{H}_{ij}(\omega)\mathbf{\Sigma}_{W_{ji}}\mathbf{H}_{ii}^*(\omega) + 
   \mathbf{H}_{ii}(\omega)\mathbf{\Sigma}_{W_{ij}}\mathbf{H}_{ij}^*(\omega)
   \label{PSDZi},
\end{equation}
where $\mathbf{H}_{ij}$ describes the transfer from $W_j$ to $Z_i$ in the frequency domain and $\mathbf{\Sigma}_{W_{ji}}=\mathbb{E}[W_{j,n}W_{i,n}^\intercal]$, the logarithmic spectral measure of the causal effect of $Z_i$ on $Z_j$ can be computed by Eq. 9 of the main paper as \cite{geweke1982measurement}
\begin{equation}
    f_{Z_j \rightarrow Z_i}(\omega)=\log \frac{|\mathbf{S}_{Z_i}(\omega)|}{|\mathbf{H}_{ii}(\omega)\mathbf{\Sigma}_{W_i}\mathbf{H}_{ii}^*(\omega)|}.
   \label{fi_j}
\end{equation}
To complete the representation of the pairwise interactions between $Z_1$ and $Z_2$, a spectral equivalent of the measure $I_{Z_1 \boldsymbol{\bigcdot} Z_2}$ given in (\ref{F1o2gauss}) can be defined as
\begin{equation}
    f_{Z_1 \boldsymbol{\bigcdot} Z_2}(\omega)=\log \frac{|\mathbf{H}_{11}(\omega)\mathbf{\Sigma}_{W_{11}}\mathbf{H}_{11}^*(\omega)||\mathbf{H}_{22}(\omega)\mathbf{\Sigma}_{W_{22}}\mathbf{H}_{22}^*(\omega)|}{|\mathbf{S}_{Z}(\omega)|},
   \label{f1o2}
\end{equation}
so as to satisfy the frequency-domain given in Eq. 10 of the main paper:
\begin{equation}
    f_{Z_1;Z_2}(\omega)=f_{Z_1 \rightarrow Z_2}(\omega) + f_{Z_2 \rightarrow Z_1}(\omega) + f_{Z_1 \boldsymbol{\bigcdot} Z_2}(\omega).
   \label{MIRexpZfreq}
\end{equation}
Note that, since the  measure (\ref{f1o2}) is defined ad-hoc to satisfy the decomposition of the total interaction, its physical meaning is not straightforward \cite{chicharro2011spectral}; as its formulation depends on both the transfer functions $\mathbf{H}_{12}$ and $\mathbf{H}_{21}$, it has been recently interpreted as a spectral measure which reflects the "mixing effects" between the directed interactions along the directions $Z_1 \rightarrow Z_2$ and $Z_2 \rightarrow Z_1$, i.e. the part of the interactions between the two processes which cannot be disentangled and assigned to one of the two causal directions \cite{pernice2021submitted}.

Importantly, all the spectral measures appearing in (\ref{MIRexpZfreq}) can be linked to the similar measures given in the time domain in (\ref{MIRexpGaussian}). In fact, it can be shown (see, e.g., \cite{chicharro2011spectral}) that the integration over the whole frequency axis of the spectral measures yields the corresponding time-domain measure:
\begin{subequations}
\begin{align}
I_{Z_1;Z_2} &= \dfrac{1}{4\pi} \int_{-\pi}^{\pi} f_{Z_1;Z_2}(\omega) \,\textrm{d}\omega \label{F12_f12},\\
T_{Z_1 \rightarrow Z_2} &= \dfrac{1}{4\pi} \int_{-\pi}^{\pi} f_{Z_1 \rightarrow Z_2}(\omega) \,\textrm{d}\omega  ,  T_{Z_2 \rightarrow Z_1} = \dfrac{1}{4\pi} \int_{-\pi}^{\pi} f_{Z_2 \rightarrow Z_1}(\omega) \,\textrm{d}\omega, \label{F1_2_f1_2}\\
I_{Z_1 \boldsymbol{\bigcdot} Z_2} &= \dfrac{1}{4\pi} \int_{-\pi}^{\pi} f_{Z_1 \boldsymbol{\bigcdot} Z_2}(\omega). \label{F1o2_f1o2}
\end{align}
\label{F_f}
\end{subequations}
The relations in (\ref{F_f}), together with the formulation of the time-domain measures (\ref{MIRexpGaussian}) in terms of MI rates, give to the spectral measures an information-theoretic meaning. In particular, the total coupling $f_{Z_1;Z_2}(\omega)$ is a non-negative quantity measuring the density of information shared between $Z_1$ and $Z_2$ at the angular frequency $\omega$. Similarly, the causal measures $f_{Z_1 \rightarrow Z_2}(\omega)$ and $f_{Z_2 \rightarrow Z_1}(\omega)$ quantify the density of the information transferred from one process to the other as a function of frequency. We note that these two measures can in general take negative values at some frequencies, although their integration over all frequencies is non-negative; negative values are possible when the off-diagonal blocks of $\mathbf{\Sigma}_W$ are non-zero, i.e. when the block process is not strictly causal. In the case of strict causality, i.e. when $\mathbf{\Sigma}_{W_{ij}}=\mathbf{\Sigma}_{W_{ji}}=0$, the non-negativity of $f_{Z_1 \rightarrow Z_2}(\omega)$ and $f_{Z_2 \rightarrow Z_1}(\omega)$ is guaranteed at each frequency \cite{chicharro2011spectral,pernice2021submitted}. On the contrary, the measure $f_{Z_1 \boldsymbol{\bigcdot} Z_2}(\omega)$ can take negative values even for strictly causal processes, as its integral over all frequencies is $I_{Z_1 \boldsymbol{\bigcdot} Z_2}=0$ when $\mathbf{\Sigma}_{W_{ij}}=\mathbf{\Sigma}_W{_{ji}}=0$ (see (\ref{F1o2gauss})) \cite{pernice2021submitted}.

Exploiting the analogy between the decompositions resulting in the time domain from (\ref{MIRexpGaussian}) and in the frequency domain from (\ref{f12}-\ref{f1o2}), we can achieve a causal decomposition of the OIR formulated for spectral functions. For instance, considering $N$ stochastic processes $\{X_1,\ldots,X_N\}$ and setting $Z_1=X^N_{-j}$ and $Z_2=X_j$, the OIR increment can be expanded in frequency as 
\begin{equation}
   \Delta_{X^N_{-j};X_j} = \dfrac{1}{4\pi} \int_{-\pi}^{\pi} \delta_{X^N_{-j};X_j}(\omega) \,\textrm{d}\omega
   \label{deltaOIRtime_freq},
\end{equation}
where the frequency-specific OIR increment is defined in analogy to (\ref{DeltaOIR}) by Eq. 12 of the main paper,
\begin{equation}
   \delta_{X^N_{-j};X_j}(\omega)=(2-N)f_{X_j;X^N_{-j}}(\omega) + \sum_{\substack{m=1 \\ m \neq j}}^{N-1}f_{X_j;X^N_{-mj}}(\omega)
   \label{deltaOIRfreq},
\end{equation}
and can be expanded through a causal decomposition similar to (\ref{deltaOIRexp}) as
\begin{equation}
   \delta_{X^N_{-j};X_j}(\omega) = \delta_{X^N_{-j} \rightarrow X_j}(\omega) + \delta_{X_j \rightarrow X^N_{-j}}(\omega) + \delta_{X^N_{-j} \bigcdot X_j}(\omega)
   \label{deltaOIRexpfreq},
\end{equation}
where the three terms on the r.h.s. of (\ref{deltaOIRexpfreq}) are obtained expanding $f_{X_j;X^N_{-j}}(\omega)$ and $f_{X_j;X^N_{-mj}}(\omega)$ in (\ref{deltaOIRfreq}) according to (\ref{MIRexpZfreq}).
Moreover, the spectral OIR increment (\ref{deltaOIRfreq}) can be used to compute recursively a frequency-domain version of the OIR, in analogy to (\ref{OIRrec}), as
\begin{equation}
   \nu_{X^N}(\omega)=\nu_{X^N_{-j}}(\omega)+\delta_{X^N_{-j};X_j}(\omega)
   \label{OIRrecfreq}.
\end{equation}
Considering (\ref{deltaOIRexpfreq}) and (\ref{OIRrecfreq}), and given (\ref{deltaOIRtime_freq}), it is easy to show that the spectral OIR and all terms of the causal decomposition of the spectral OIR increment satisfy individually the spectral integration property, i.e. the average over all frequencies of each of these spectral functions yields the corresponding information-theoretic function. Therefore, the spectral versions of the high-order interaction measures defined in this section can be meaningfully interpreted as densities of the synergistic/redundant character of the information shared between multiple stochastic processes.

\section*{Theoretical examples}
\setcounter{subsection}{0}
\subsection{Simulation 1}
The first simulation considers three scalar Gaussian processes $X_1,X_2,X_3$ whose dynamics and interactions are defined by the trivariate VAR model:
\begin{subequations}
\begin{align}
X_{1,n} &= \sum_{k=1}^{4} a_{11,k} X_{1,n-k}+ U_{1,n}\\
X_{2,n} &= \sum_{k=1}^{q+1} a_{21,k} X_{1,n-k} + a_{22,1} X_{2,n-1} + a_{22,2} X_{2,n-2}  + U_{2,n}\\
X_{3,n} &= \sum_{k=1}^{q+1} a_{31,k} X_{1,n-k} + a_{32} X_{2,n-1} + U_{3,n}
\end{align}
\label{simu1_eq}
\end{subequations}
In (\ref{simu1_eq}), $U_1$, $U_2$, and $U_3$ are white uncorrelated Gaussian noise processes with zero mean and variance set respectively to $\sigma^2_{U_1}=2$, $\sigma^2_{U_2}=0.5$, and $\sigma^2_{U_3}=2$.
The coefficients determining self-dependencies (i.e., $a_{ii,k}, i=1,2$) are set placing two pairs of complex-conjugate poles in the complex plane (with modulus $\rho_{11}=0.85$ and $\rho_{12}=0.85$ and phases $\phi_{11} =2 \pi 0.1$ rad and $\phi_{12} =2 \pi 0.35$ rad) to generate autonomous oscillations at the frequencies $\sim 0.1$ Hz and $\sim 0.35$ Hz for $X_1$, and placing one pair of complex conjugate poles in the complex plane (with modulus $\rho_2=0.7$ and phase $\phi_2 =2 \pi 0.1$ rad) to generate an autonomous oscillation at the frequency $\sim 0.1$ Hz for $X_2$ (here, we denote frequencies in Hz assuming sampling frequency $f_s=1$).
The coefficients determining causal dependencies (i.e., $a_{ij,k}, i,j=1,2,3, i \neq j$) are set as follows: $a_{21,k}=0.4b_{1,k}$, where $b_{1,k}$ are the coefficients of a low-pass FIR filter of order $q=20$, with cutoff frequency at 0.2 Hz;
$a_{31,k}=0.6b_{2,k}$, where $b_{2,k}$ are the coefficients of a high-pass FIR filter of order $q=20$, with cutoff frequency at 0.2 Hz; $a_32=1$ to realize an all-pass filter.

This simulation is implemented by the script \texttt{test\textunderscore oir\textunderscore simu1.m} of the OIR Matlab toolbox, which produces the results shown in Fig. 1 of the main paper.

\subsection{Simulation 2}
The second simulation considers ten scalar Gaussian processes whose dynamics and interactions are defined by the 10-variate VAR model:
\begin{subequations}
\begin{align}
Y_{1,n} &= 2\rho_1 \cos (2\pi f_1) Y_{1,n-1} -\rho_1^2 Y_{1,n-2} + U_{1,n}\\
Y_{2,n} &= 0.5 Y_{1,n-1} + U_{2,n}\\
Y_{3,n} &= 0.5 Y_{2,n-1} + U_{3,n}\\
Y_{4,n} &= -0.5 Y_{1,n-2} + 0.2 Y_{3,n-1} + 0.5 Y_{10,n-1} + U_{4,n}\\
Y_{5,n} &= 2\rho_1 \cos (2\pi f_1) Y_{5,n-1} -\rho_1^2 Y_{5,n-2} + U_{5,n}\\
Y_{6,n} &= 0.3 Y_{7,n-2} + U_{6,n}\\
Y_{7,n} &= 2\rho_1 \cos (2\pi f_1) Y_{7,n-1} -\rho_1^2 Y_{7,n-2} + 0.3 Y_{6,n-1} + U_{7,n}\\
Y_{8,n} &= 2\rho_2 \cos (2\pi f_2) Y_{8,n-1} -\rho_2^2 Y_{8,n-2} + 0.4 Y_{2,n-2} + 0.3 Y_{3,n-1} -0.4 Y_{5,n-1} + 0.3 Y_{7,n-1} + U_{8,n}\\
Y_{9,n} &= 0.7 Y_{8,n-1} - 0.2 Y_{10,n-2} + U_{9,n}\\
Y_{10,n} &= 0.4 Y_{9,n-1} + U_{10,n}
\end{align}
\label{simu2_eq}
\end{subequations}
In (\ref{simu2_eq}), the innovation processes $U_1, \ldots U_{10}$ are white uncorrelated Gaussian noise processes with zero mean and unit variance.
The coefficients determining self-dependencies (i.e., $a_{ii,k}, i=1,2$) are set placing one pair of complex conjugate poles in the complex plane to generate an autonomous oscillation at the frequency $\sim 10$ Hz for $Y_1$, $Y_5$ and $Y_7$ (pole modulus $\rho_1=0.9$, pole frequency $f_1=10/f_s=0.1$ Hz), and  at the frequency $\sim 25$ Hz for $Y_8$ (pole modulus $\rho_2=0.8$, pole frequency $f_1=25/f_s=0.1$ Hz), with simulated sampling frequency $f_s=100$ Hz.
The coefficients determining causal dependencies are set with the values indicated in (\ref{simu2_eq}), to obtain directed dependencies between pairs of processes as depicted in Fig. 3a of the main paper.
The analysis of HOIs is then performed after grouping the original 10 processes into 5 blocks organized as follows: $X_1=\{Y_1,Y_2,Y_3,Y_4\}$; $X_2=\{Y_5\}$; $X_3=\{Y_6,Y_7\}$; $X_4=\{Y_8\}$; $X_5=\{Y_9,Y_{10}\}$.

This simulation is implemented by the script \texttt{test\textunderscore oir\textunderscore simu2.m} of the OIR Matlab toolbox, which produces the results shown in Fig. 3 of the main paper.

\section*{Cardiovascular and respiratory interactions during paced breathing}
\setcounter{subsection}{0}
\subsection{Data acquisition and experimental protocol}
The analyzed data belong to an historical database of cardiovascular and respiratory time series measured during a protocol of paced breathing \cite{porta2000information,porta2011accounting}.
The signals were recorded with a sampling frequency of $f_s=300$ Hz on a control group composed of 19 healthy subjects (11 females, 8 males, age: 27-35, median: 31 years). Data consisted in electrocardiographic (ECG, lead II), noninvasive arterial blood pressure (BP) (Finapres 2300, Ohmeda, Englewood, CO), and respiratory flow (RF) (nasal thermistor by Marazza, Monza, Italy) signals. The experimental protocol included four different sessions, with subjects laying in the supine position normally breathing (spontaneous breathing, SB) or following a controlled metronome breathing at 10, 15 and 20 breaths/min (CB10, CB15 and CB20, respectively). The period of spontaneous respiration was always carried out first, followed by the controlled breathing sessions instead performed in a random order. The data of one subject were excluded from the analyses, since the SAP time series, extracted from the corresponding BP signal, was not readable during the condition CB20. The total number of analysed subjects was then reduced to 18.
\subsection{Data pre-processing}
Three time series were extracted from ECG, AP and RF signals on a beat-to-beat basis as follows: (i) the heart period (HP) series was extracted as the sequence of the temporal distances between consecutive R peaks (R-R intervals) of the ECG signal; (ii) the systolic arterial pressure (SAP) series was obtained as the sequences of the maximum values of the BP signal measured within each detected R-R interval; and (iii) the respiration (RESP) series was extracted as the sequence of RF values sampled at the onset of each detected R-R interval. Further information about signal acquisition and series extraction can be found in \cite{porta2000information,porta2011accounting}. For each subject and experimental condition, stationary segments of $N=256$ points were selected through a visual inspection. Before the analysis, the series were linearly detrended using a zero-phase autoregressive (AR) high-pass filter with a cut-off frequency equal to 0.0156 $f_s$, and then normalized to zero mean \cite{nollo2000synchronization}.

\subsection{Data analysis}
The analysis was performed computing, in both time and spectral domains,  the OIR increments obtained adding the HP series ($X_3$) to the bivariate network formed by the RESP ($X_1$) and SAP ($X_2$) series.
Specifically, we identified a VAR model fitting the vector process comprising the three time series of interest ($X_1$, $X_2$, $X_3$) by using least squares identification. Model order selection was first performed using the Akaike criterion, setting the maximum lag to 14. Then, model orders were manually adjusted by visual inspection in the range 3-8.
The OIR toolbox was then used to compute, in the frequency domain, the OIR increment relevant to the addition of HP to $\{$SAP, RESP$\}$ ($\delta_{X_3;X_1,X_2}(\omega)$), the causal OIR increment from HP to $\{$SAP, RESP$\}$ ($\Delta_{X_3{\rightarrow}X_1,X_2}(\omega)$), the causal OIR increment from $\{$SAP, RESP$\}$ to HP ($\Delta_{X_1,X_2{\rightarrow}X_3}(\omega)$) and the instantaneous OIR increment ($\Delta_{X_3{\cdot}X_1,X_2}(\omega)$). 
The spectral profiles were integrated within the Low Frequency (LF) and High Frequency (HF) bands defined as follows: regarding the LF band, we considered the range [0.04-0.12 Hz]; regarding the HF band, we first identified the respiratory peak within the range [0.12-0.4] Hz for each subject, and then selected the range around the peak with a width of $\pm 0.04$ Hz.

\subsection{Statistical analysis}
The statistical significance of the distributions obtained for the measures was performed using non-parametric tests, given the small sample size and since the assumption of normality was rejected for most distributions using the Anderson-Darling test. The non-parametric one-way Friedman test was employed to assess the statistical significance of the differences of the median of the distributions, followed by a post-hoc Wilcoxon test with Bonferroni-Holm correction for multiple comparison ($n=3$) to assess the difference between each distribution in a controlled breathing condition versus the spontaneous breathing reference (i.e. C10 vs SB, C15 vs SB, C20 vs SB). All the statistical tests were carried out with 5\% significance level.

\subsection{Results}
The OIR decomposition analysis of cardiovascular time series is implemented by the script \texttt{test\textunderscore oir\textunderscore HPRESPSAP.m} of the OIR Matlab toolbox, which produces for one representative subject the spectral profiles shown in Fig. 4a of the main paper.

\begin{figure} [t]
    \centering
    \includegraphics[scale=0.52]{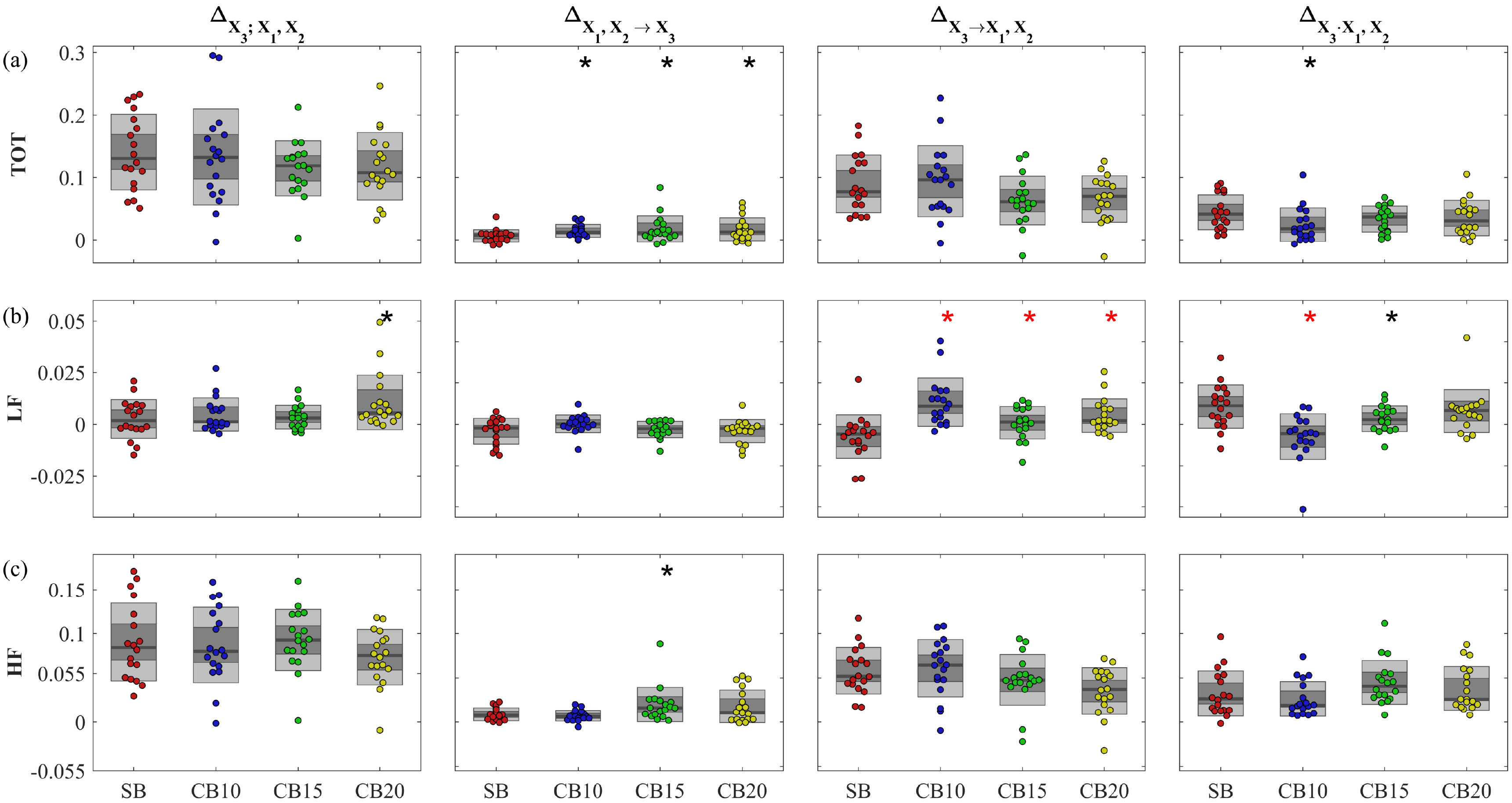}
    \caption{OIR decomposition of cardiovascular and cardiorespiratory interactions computed during spontaneous breathing (SB) and controlled breathing (CB) at 10 (CB10), 15 (CB15) and 20 (CB20) breaths/min. (a), Time-domain, (b) LF (0.04-0.12 Hz) and (c) HF ($f_{RESP}\pm0.04$ Hz) values of the total OIR increment ($\Delta_{X_3;X_1,X_2}$) obtained with the addition of HP to $\{$SAP, RESP$\}$, of the causal OIR increments from $\{$SAP, RESP$\}$ to HP ($\Delta_{X_1,X_2{\rightarrow}X_3}$) and from HP to $\{$SAP, RESP$\}$ ($\Delta_{X_3{\rightarrow}X_1,X_2}$), and the instantaneous OIR increment ($\Delta_{X_3{\cdot}X_1,X_2}$). Asterisks denote statistically significant difference between the CB condition compared with SB (Wilcoxon signed-rank test: black, uncorrected; red, Bonferroni-Holm correction for multiple comparisons).}
    \label{fig_applCardio_OIR_supp}
\end{figure}

Fig. \ref{fig_applCardio_OIR_supp} reports the distributions across subjects (boxplots and individual values) of the OIR increments obtained adding HP to $\{$SAP, RESP$\}$, as well as of the two causal terms and the instantaneous term of the decomposition of such increments, computed in the four experimental conditions and integrated over all frequencies as well as within the selected LF and HF bands. 

\section*{Neural interactions from ECoG signals in the
anesthetized macaque monkey}
\setcounter{subsection}{0}
\subsection{Data acquisition and experimental protocol}
The dataset used in this study can be downloaded from \href{http://neurotycho.org/expdatalist/listview?task=45}{http://neurotycho.org/expdatalist/listview?task=45}; the analyzed data are relevant to the macaque monkey named Su. During the experiment analyzed in our work, the monkey was seated in a primate chair with both arms and head movement restricted and eyes covered to avoid evoking visual response during the experimental period. We considered two experimental conditions, relevant to a resting state before (REST) and after (ANES) the injection of a cocktail of anesthetics consisting of ketamine hydrochloride ($1.15$ ml ($8.8$mg/Kg)) and Medetomidine ($0.35$ ml ($0.05$ mg/Kg)).
All the experimental and surgical procedures were performed in a previous study according with the experimental protocol (No. H24-2-203(4)) approved by the RIKEN ethics committee \cite{yanagawa2013large}. 

The complete description of the surgical implantation of the ECoG electrodes can be found in \cite{nagasaka2011multidimensional}. The acquired data consisted of 128 ECoG signals recorded with a sampling frequency of 1 kHz with electrodes placed in pairs with an inter-electrode distance of 5 mm to cover the entire left hemisphere of the brain. 

 \subsection{Data pre-processing}
 We considered two five-minutes recording sessions during the REST and ANES conditions.
 The ten bipolar ECoG signals selected for the analysis (Figure 5a of the main paper) were band-pass filtered between 0.5 and 200 Hz to remove slow and fast components in the power spectrum (zero-phase Butterworth filter; notch filter 49-51 Hz with a slope of 48 dB/oct in the transition band), downsampled to $f_s=250$ Hz, epoched to extract $\sim 160$ trials lasting 2 sec for each condition, and finally normalized to zero mean and unit variance within each trial.
 
\subsection{Data analysis}
The analysis of high-order interactions was performed starting from the VAR parameters of the model fitting each 2-sec trial. Considering the $M=5$ blocks of time series acquired from the 10 pairs of bipolar electrodes analyzed, the spectral OIR was computed for all multiplets of order $N=3,4,5$.
Time-domain OIR values ($\Omega$) were then obtained integrating the spectral measures $\nu(f)$ within the $\delta$ (0.2-3 Hz), $\theta$ (4-7 Hz), $\alpha$ (8-12 Hz), $\beta$ (12-30 Hz) and $\gamma$ (31-70 Hz) frequency bands, as well as cumulatively between 0 and 70 Hz (T).

\subsection{Statistical analysis}
Since we were only interested in the difference between the two experimental conditions (REST vs ANES), irrespective of the analyzed multiplet of time series, we performed, for each multiplet and for each interval of integration, a Wilcoxon signed rank test with significance level ($\alpha$) equal to 0.05, followed by Bonferroni correction for multiple comparisons (six multiple comparison were considered, i.e. those between REST and ANES relevant to the OIR measure obtained within the $\delta$, $\theta$, $\alpha$, $\beta$ and $\gamma$ bands, plus the whole-band time domain measure).

\subsection{Results}
The OIR decomposition analysis for the ECoG time series is implemented by the script \texttt{test\textunderscore oir\textunderscore ECoG.m} of the OIR Matlab toolbox, which produces for one representative 2-sec trial the spectral profiles shown in Fig. 5b of the main paper.

\begin{figure}[t]
    \centering
    \includegraphics{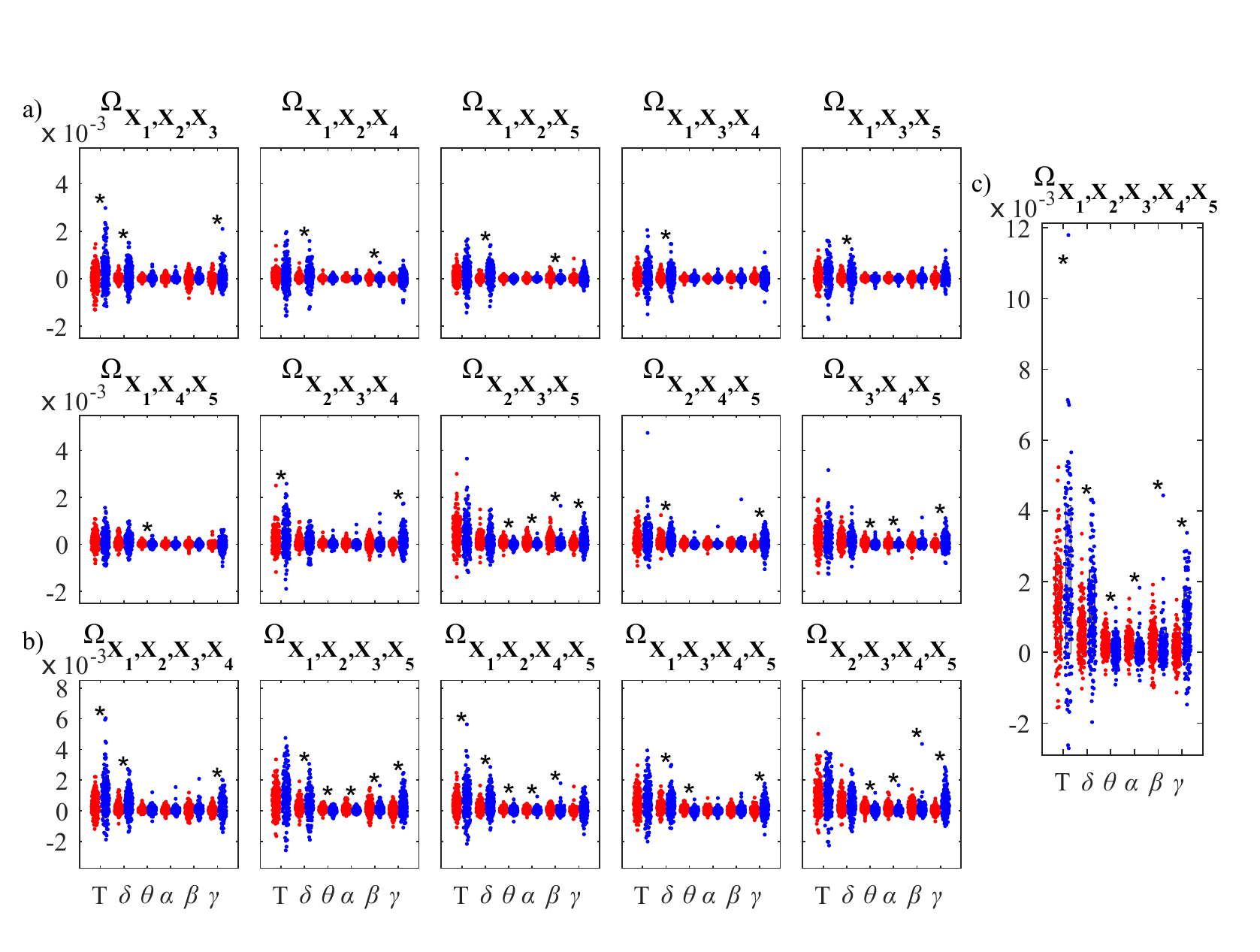}
    \caption{OIR analysis of neurophysiological interactions in the anesthetized monkey named Su. The plots depict the values (distribution across $\sim 160$  trials) of the time-domain measure of HOIs obtained by integrating the spectral OIR over the whole frequency axis (T) or within the $\delta$,$\theta$,$\alpha$,$\beta$ and $\gamma$ bands. Results are reported for all multiplets of order 3 (a), all multiplets of order 4 (a), and the multiplet of order 5 including all time series (c). Asterisks denote statistically significant difference between REST (red dots) and ANES (blue dots) obtained after Wilcoxon singed-rank test with Bonferroni correction ($p<\alpha_c/6)$).} \label{fig_applECOG_OIR_supp}
\end{figure}

Fig. \ref{fig_applECOG_OIR_supp} reports the distributions across subjects (boxplots and individual values) of the OIR, computed in the two experimental conditions (REST, ANES) and integrated over all frequencies (T) as well as within the five selected frequency bands ( $\delta$, $\theta$, $\alpha$, $\beta$ and $\gamma$). 

\end{document}